\documentclass[%
 reprint,
superscriptaddress,
 amsmath,amssymb,
 aps,
 pre,
]{revtex4-1}

\usepackage{graphicx}
\usepackage{dcolumn}
\usepackage{bm}

\begin{document}


\title{Buoyancy Effects on the Scaling Characteristics of Atmospheric Boundary Layer Wind Fields in the Mesoscale Range}

\author{V. P. Kiliyanpilakkil}
\author{S. Basu}%
\email{sukanta\_basu@ncsu.edu}
\affiliation{%
 Department of Marine, Earth, and Atmospheric Sciences, North Carolina State University, Raleigh, NC 27695, USA}
\author{A. Ruiz-Columbi\'{e}}
\affiliation{National Wind Institute, Texas Tech University, Lubbock, TX 79409, USA}
\author{G. Araya}
\affiliation{Department of Mechanical Engineering, University of Puerto Rico, Mayaguez,
Puerto Rico 00681-9045, USA}
\author{L. Castillo}
\affiliation{Department of Mechanical Engineering, Texas Tech University, Lubbock, TX 79409, USA}
\author{B. Hirth}
\author{W. Burgett}
\affiliation{National Wind Institute, Texas Tech University, Lubbock, TX 79409, USA}

\date{\today}

\begin{abstract}
We have analyzed long-term wind speed time-series from five field sites up to a height of 300 m from the ground. Structure function-based scaling analysis has revealed that the scaling exponents in the mesoscale regime systematically depend on height. This anomalous behavior is likely caused by the buoyancy effects. In the framework of the extended self-similarity, the relative scaling exponents portray quasi-universal behavior.  
\end{abstract}

\pacs{Valid PACS appear here}
\maketitle

\section{Introduction}

The atmospheric boundary layer (ABL) spans the lowest few hundred meters \footnote{During nighttime, the depth of the boundary layer ($h$) can be very shallow; $h \ll 100$ m. In contrast, during the daytime (over land), $h$ can be on the order of 2--3 km.} of the earth's atmosphere and intensively exchanges mass (e.g., moisture), momentum, and heat with the underlying surface \cite{stul88,garr92}. Wind fields in this turbulent layer play major roles in a wide range of industrial (e.g., stack gas dispersion, wind energy generation), biological (e.g., evapotranspiration, pollen transport, migrations of birds and insects), and natural (e.g., soil erosion, transport, and deposition) activities and processes. Thus, it is not surprising that numerous studies have been conducted over the past decades for the multiscale characterizations of the ABL wind fields. Diverse methodologies with varying degrees of complexities, ranging from the traditional spectral methods  \cite{van1957power,kolesnikova1968spectra,lyons1975mesoscale,ishida1990seasonal,
larsen2011case} to the new-generation multifractal approaches  \cite{chambers1984atmospheric,sreenivasan1993update,praskovsky1997comprehensive,
kurien2000scaling,basu2007estimating,morales2012characterization}, have been utilized. 

To this date, the majority of the studies dealing with the identification of anomalous scaling (sometimes even multifractality \footnote{Quite a few present-day authors inappropriately use the terms anomalous scaling and multifractality interchangeably. For specific cases, the multifractal formalism is a plausible way of explaining anomalous scaling behavior. However, it is not always applicable.}) in the ABL wind field focused on the microscale range (time-scale of tenths of a second to tens of minutes). In contrast, only a handful of studies \cite{lauren1999characterisation,
lauren2001analysis,govindan2004long,kavasseri2005multifractal,koccak2009examination,
muzy2010intermittency,baile2010spatial,telesca2011analysis,liu2013cascade} delved into characterizing the mesoscale range (approximately, sub-hourly to sub-daily time-scale). They analyzed observational datasets from field sites around the world (e.g., China, France, New Zealand, Italy, the Netherlands, Turkey, and the USA). Even though a few of these studies did not use wind data with adequate temporal resolution or sample size, the evidences of anomalous scaling in the mesoscale range were beyond any doubt. Most remarkably, Muzy et al. \cite{muzy2010intermittency} and Ba{\"\i}le and Muzy \cite{baile2010spatial} found the intermittent nature of mesoscale wind fluctuations to be similar to its microscale counterpart (encompassing the inertial-range of turbulence). Similar conclusions were also recently drawn by Liu and Hu \cite{liu2013cascade}.

Owing to the dearth of long-term, high-quality upper ABL data, almost all the aforementioned mesoscale wind characterization studies focused on the near-surface (around 10--20 m from the ground) region. The only exception being the study by Telesca and Lovallo \cite{telesca2011analysis}. They analyzed multi-year sodar-based wind data from various heights (50 m to 213 m) above ground level (AGL). They used the Multifractal Detrended Fluctuation Analysis and the Fisher-Shannon Information Plane approaches to detect any signature of multifractality in wind speed time-series. Interestingly, they found the scaling exponents to be strongly dependent on height. However, no physical explanation was provided. It is, however, plausible to speculate that the buoyancy effects are at the root of this height-dependency trait.  

In the literature, it is well-known that the shear production of turbulence overpowers the buoyancy effects near the surface. However, buoyancy forcing becomes increasingly dominant as one moves away from the surface \cite{stul88,garr92}. However, it is not known whether the buoyancy forcing modulates the anomalous scaling behavior of wind speed in the mesoscale range. Therefore, we decided to address this intriguing issue in the present study. 

\section{Description of Sites and Datasets}

We make use of long-term wind datasets from several field sites with diverse geographical and climatological conditions (Table~\ref{tab:table1}). These datasets are measured with the aid of different types of research-grade instruments (e.g., cup anemometers, sodars). They all have a common averaging time of 10 min.
More importantly, they all span the lower part of the ABL and not just the near-surface region. 

FINO 1 is an offshore platform in the North Sea \cite{neumann2003erection,turk2008wind,ernst2012investigation}. It consists of a 100 m tall meteorological tower equipped with wind speed measurement sensors (cup anemometers) at heights of 33 m, 40 m, 50 m, 60 m, 70 m, 80 m, 90 m, and 100 m. 
A total of 91 months of wind speed data collected over a period of nine years (2004--2012) are utilized in the present study. 

Over the past four decades, observational data from the Cabauw (the Netherlands) meteorological tower have been used in various ABL studies \cite{nieuwstadt1978computation,nieuwstadt1984turbulent,beljaars1991flux,verkaik2007wind}.
We use 170 months of wind speed data (collected during the years 2001--2015) measured by propeller vanes at heights of 10 m, 20 m, 40 m, 80 m, 140 m, and 200 m. We would like to point out that even though the landscape at Cabauw is quite flat and open (grassland), the existence of wind breaks and scattered villages cause significant disturbances in the near-surface region \cite{verkaik2007wind}. The impact of this non-equilibrium behavior on scaling characteristics will be noted later in this paper.

Recently, the West Texas Mesonet (WTM) has installed sodars (manufacturer: Scintec, model: MFAS) at San Angelo, Midland, and Reese Technology Center (RTC), Lubbock (the USA). The typical vertical range of these sodars is from 30 m to approximately 300 m AGL. They have a vertical resolution of 10 m. From the past two years (2013--2014), we selected 8 months of wind speed data for scaling analysis.  

All the aforementioned datasets contain variable amount of missing data. This data-loss problem is more severe for the sodars. Since a sodar is an active ground based acoustic remote sensing instrument, it suffers from signal attenuation at higher altitudes \cite{bradley2008atmospheric}. To account for the data-loss problem in an objective manner across the sites, we perform the following data pre-processing procedures. First, instead of analyzing an entire time-series (with sporadic gaps) from any site, we split it into several monthly time-series (each containing  about 4300 samples). We discard a specific month's data from further analysis if any of the vertical levels contain more than 20\% of missing samples. By performing this simple data exclusion strategy, we ensure that for a given site, all the tower/sodar levels have more-or-less (within 20\%) the same amount of samples. In the case of the WTM sodar data, we have added an additional constraint. We consider only the months during which all the three sodars collected wind data simultaneously. 

Prior to scaling analysis, we normalize (zero mean, unit variance) each monthly time-series. Various orders of structure functions (defined below) are computed based on each normalized monthly time-series and then averaged over the different months. The normalization procedure aids in the visual detection of any collapse of computed statistics from various heights (AGL). 

\begin{table}[t]
\caption{\label{tab:table1}%
Description of measurement sites
}
\begin{ruledtabular}
\begin{tabular}{lccc}
\textrm{Site}& \textrm{Elevation (m)\footnote{Mean sea level}}& \textrm{Location} & \textrm{\# Months}\\
\colrule
FINO-1 & 0 & 54.01$^\circ$ N, & 91\\
       &   &  6.59$^\circ$ E  & \\
Cabauw & -0.7 & 51.97$^\circ$ N, & 170\\
       &   &  4.93$^\circ$ E  & \\
San Angelo & 597.4 & 31.54$^\circ$ N,  & 8 \\
           &       & 100.51$^\circ$ W  & \\ 
Midland    & 874.8 & 31.95$^\circ$ N,  & 8 \\
           &       & 102.21$^\circ$ W  & \\
RTC        & 1015.6 & 33.60$^\circ$ N,  & 8 \\
           &        & 102.04$^\circ$ W & \\
\end{tabular}
\end{ruledtabular}
\end{table}

\section{Structure Function Analysis}

In   the  turbulence  literature,   the  scaling   exponent  spectrum,
$\zeta_p$, is defined as \cite{frisch1995turbulence,bohr1998dynamical}:
\begin{equation}
S_p(r) = \langle |\Delta u|^p \rangle \sim r^{\zeta_p}
\end{equation}
where $S_p(r)$  is the so-called  $p$-th order structure  function. The  angular bracket denotes spatial averaging  and $r$  is a separation distance that varies  within a specific scaling range (e.g., inertial-range). For time-series analysis (where Taylor's hypothesis is inapplicable), the usage of time increment, $\Delta t$ (in lieu of $r$), and temporal averaging is customary. This approach is followed here. 

According to Kolmogorov's celebrated 1941 hypothesis (K-41; \cite{kolmogorov1941local}), $\zeta_p = p/3$ in the isotropic inertial-range of turbulence. In the buoyancy-range, for the velocity field, the hypothesis of Bolgiano \cite{bolgiano1959turbulent,bolgiano1962structure,monin1975} leads to: $\zeta_p = \frac{3}{5}p$. Over the years, several laboratory and numerical studies \cite{benzi1994scaling,niemela2000turbulent,boffetta2012bolgiano} have corroborated the existence of Bolgiano scaling in different types of convection. Its presence has also been indicated in the scaling of near-surface temperature field \cite{aivalis2002temperature} and vertical wind speed profiles \cite{lovejoy2007isotropic,lovejoy2013weather}. However, to the best of our knowledge, this scaling has never been reported for the ABL (horizontal) wind field. Forty years ago, Monin and Yaglom (pp. 393 of \cite{monin1975}) wrote: ``It is therefore probable that the formulas given above [in the context of Bolgiano scaling] will be valid beginning with heights of the order of 100 m [above the surface]. The verification of this conclusion will require special observations which one hopes will be carried out in the future"\footnote{The text inside the parentheses are made by the authors of the present paper and not by \cite{monin1975}}. We are fortunate to have access to such `special observations', and thus, will be able to address a few unresolved scaling issues of wind speed in the mesoscale range.

\begin{figure*}[!ht]
\includegraphics[width=2.3in]{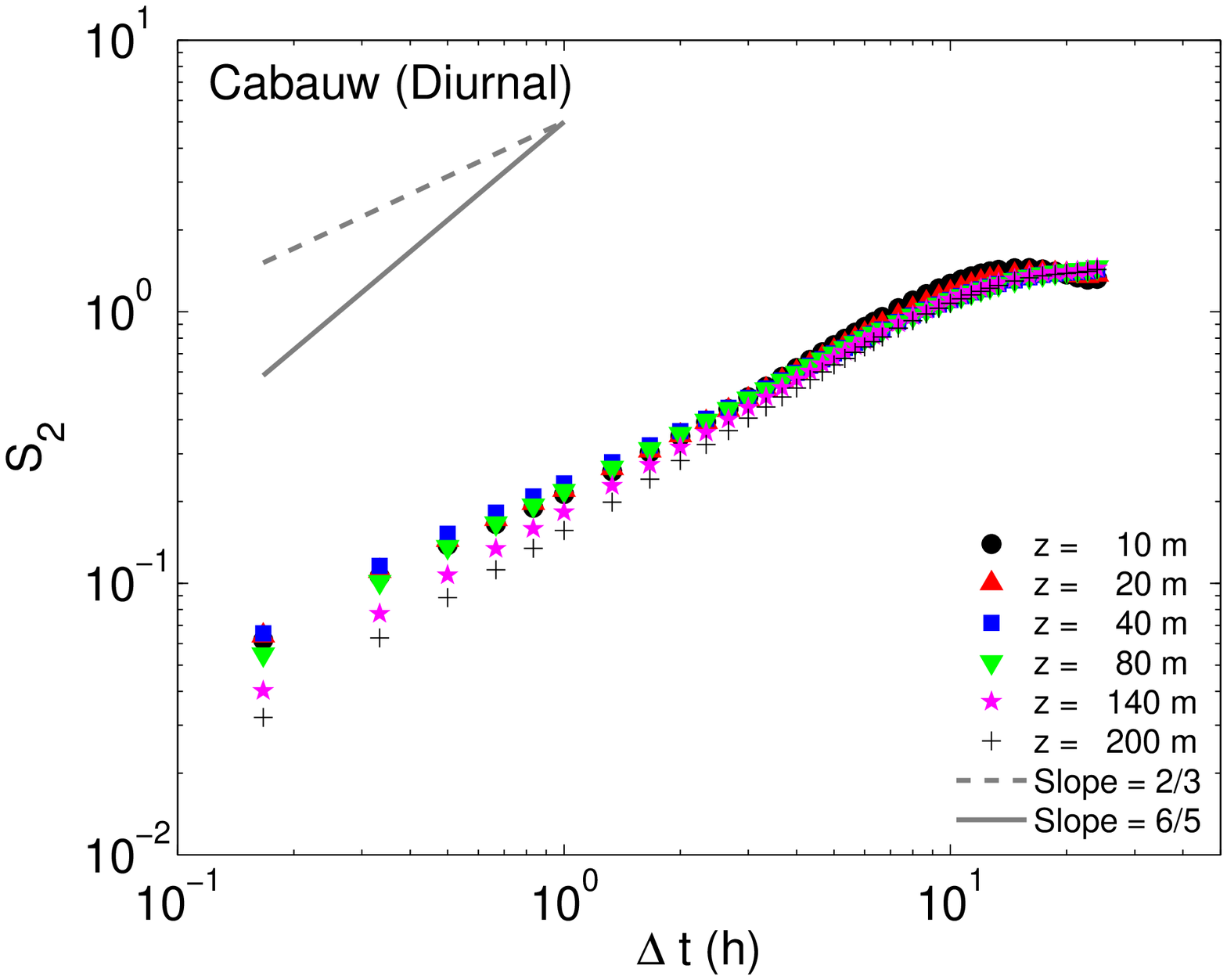}
\includegraphics[width=2.3in]{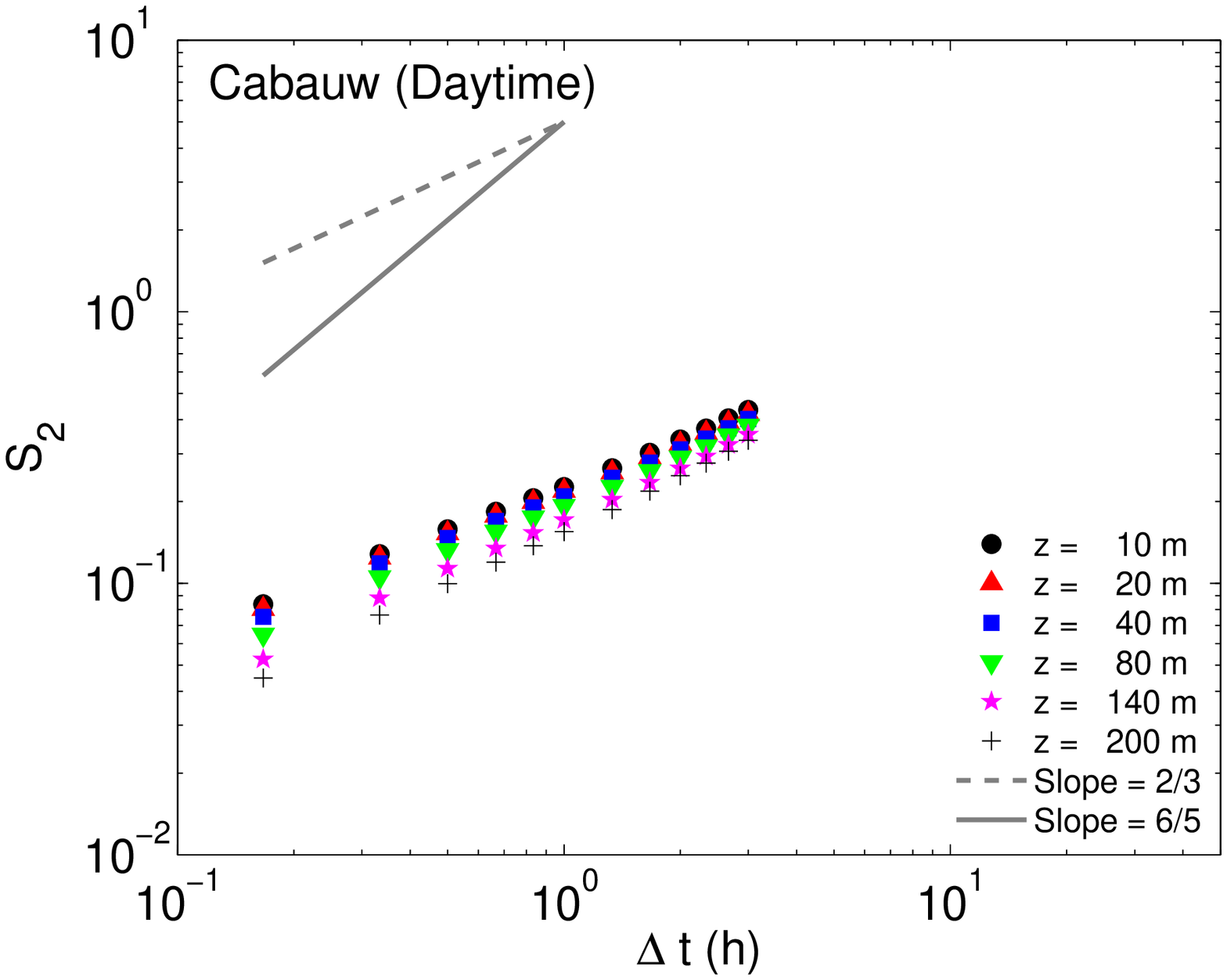}
\includegraphics[width=2.3in]{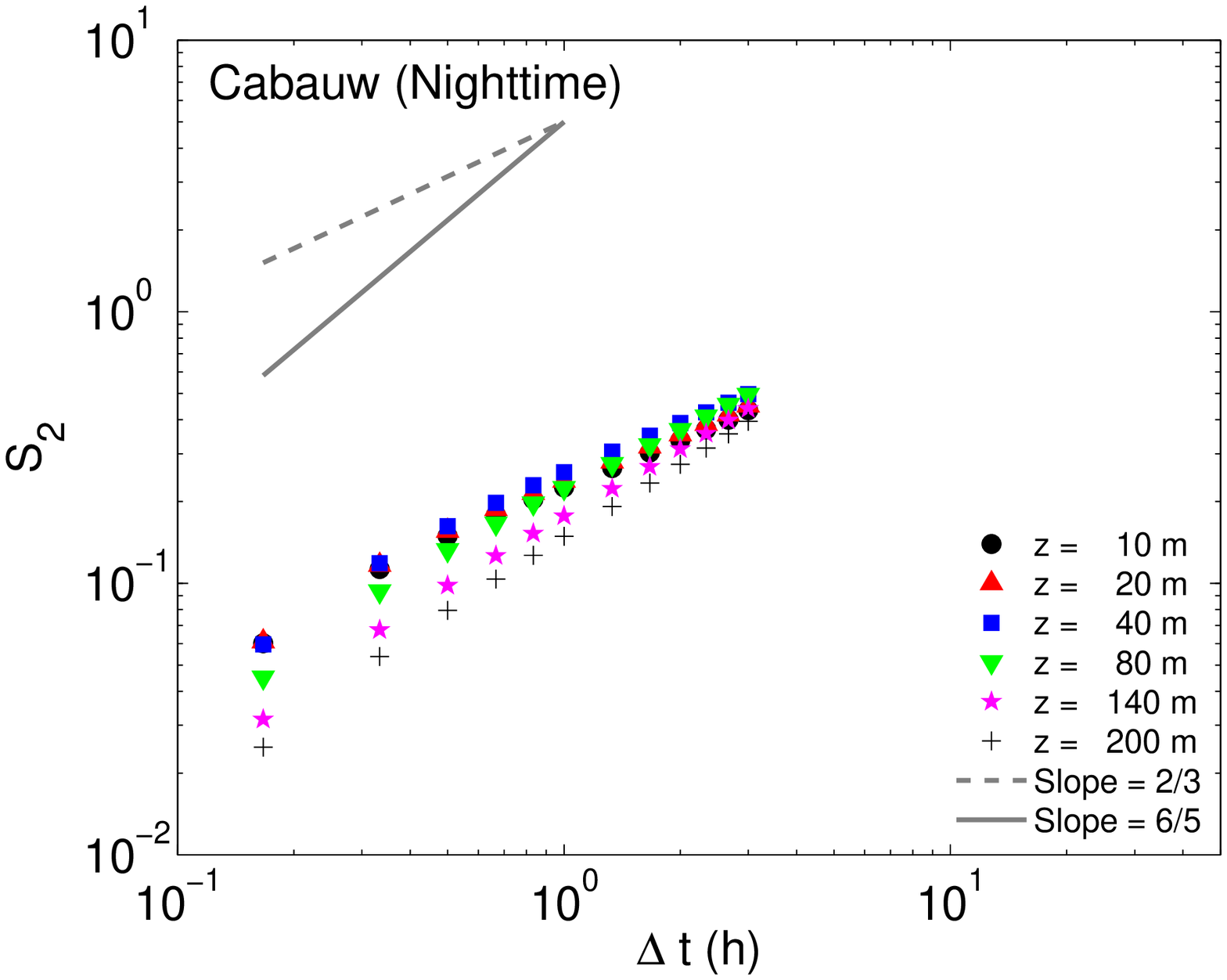}\\
\includegraphics[width=2.3in]{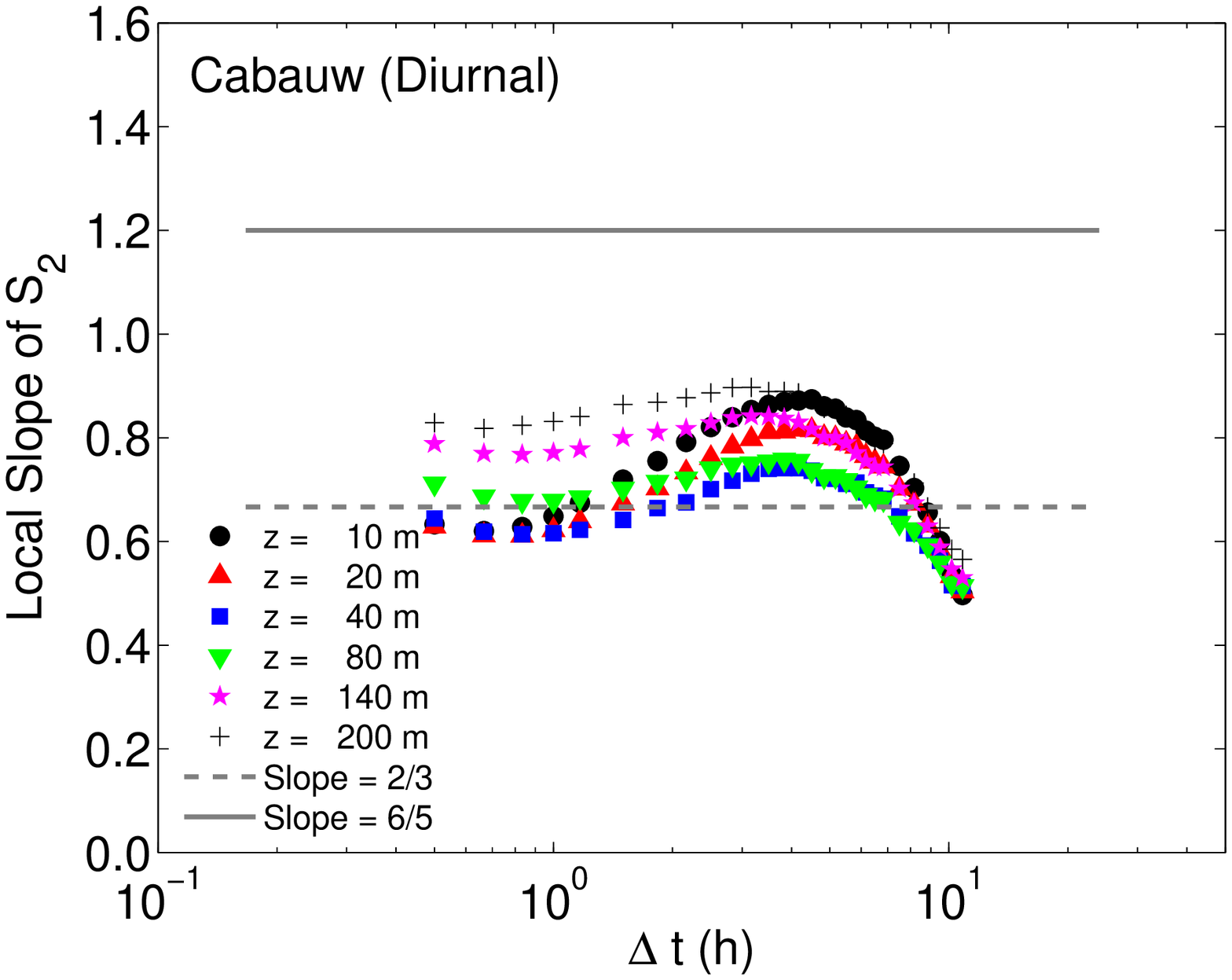}
\includegraphics[width=2.3in]{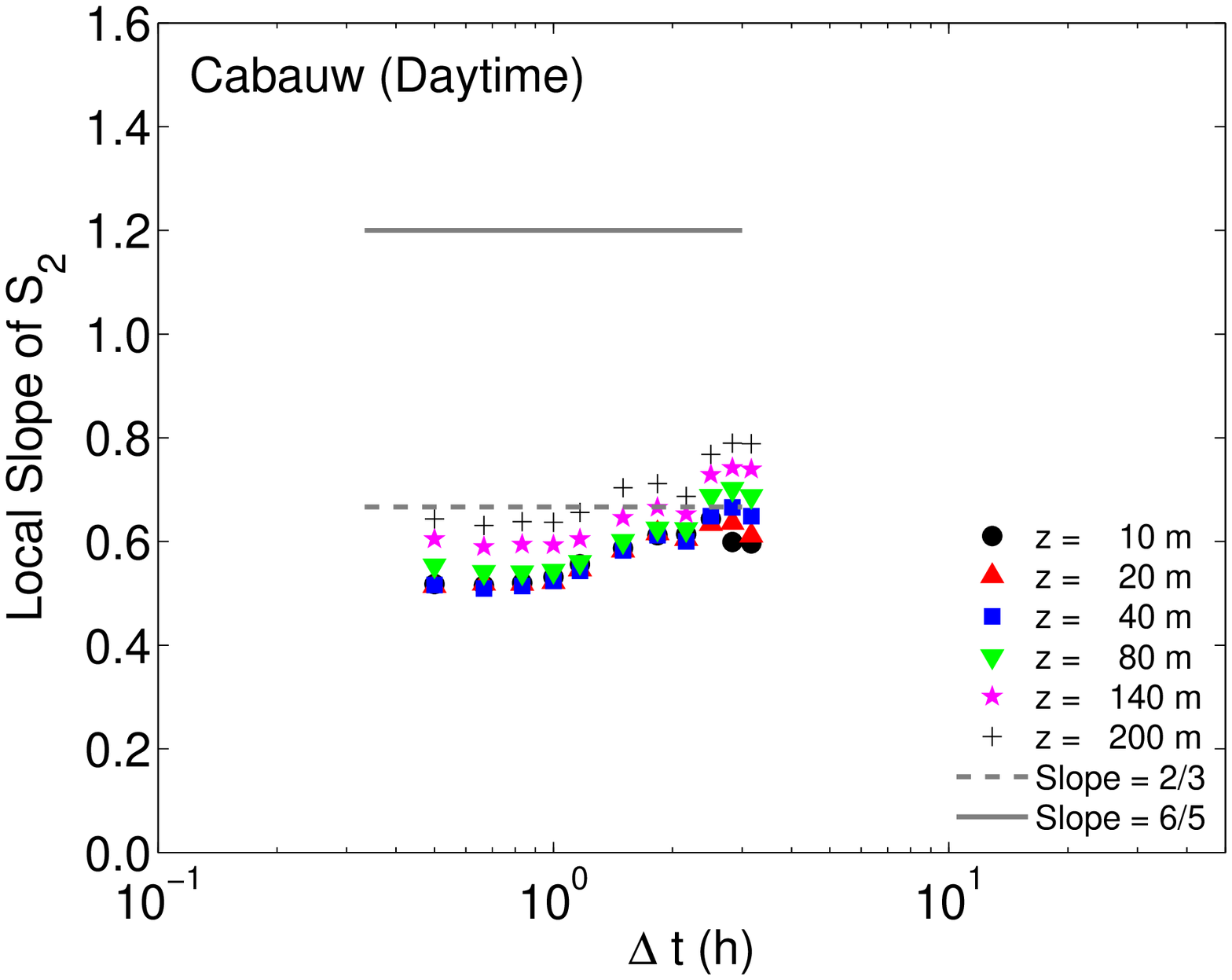}
\includegraphics[width=2.3in]{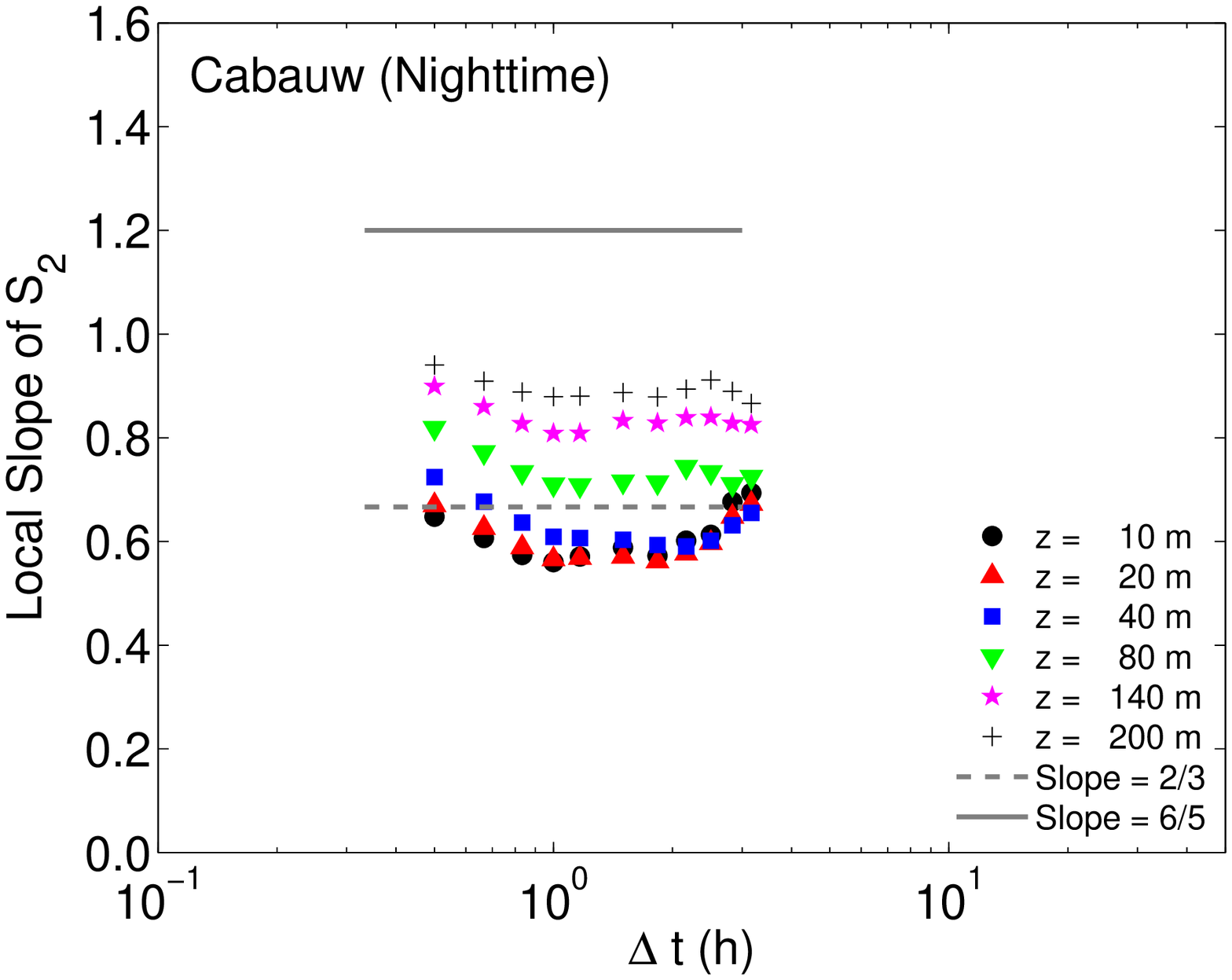}\\
\caption{\label{SF2-Cabauw} (Color Online) The second-order structure functions (top panel) and their corresponding local slopes (bottom panel) based on wind speed observations from the Cabauw tower. The left, middle, and right panels represent diurnal, daytime (9--15 UTC), and nighttime (21--3 UTC) data, respectively. The dashed and solid lines represent slopes corresponding to Kolmogorov's and Bolgiano's hypotheses, respectively.}
\end{figure*}

\begin{figure*}[t]
\includegraphics[width=2.3in]{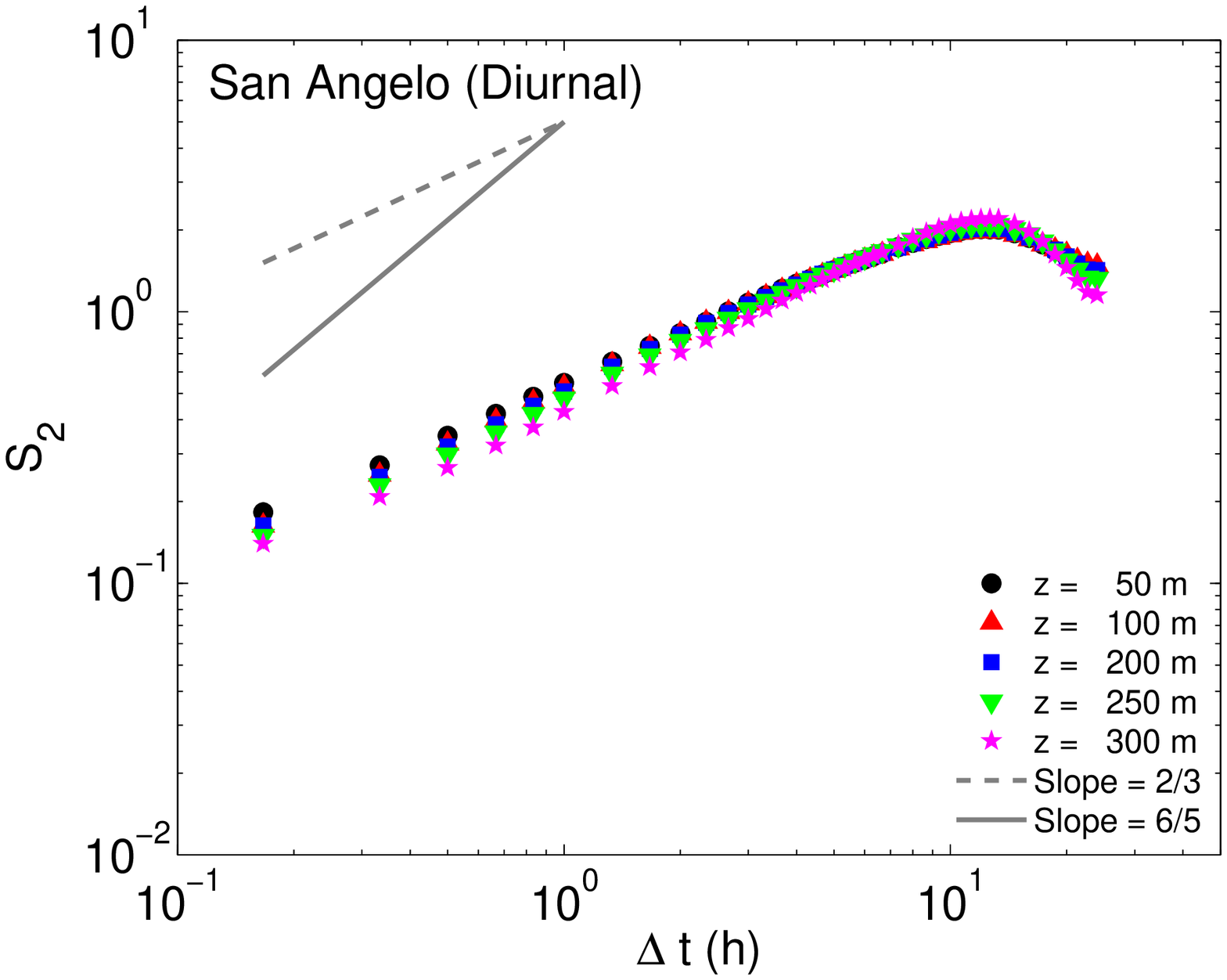}
\includegraphics[width=2.3in]{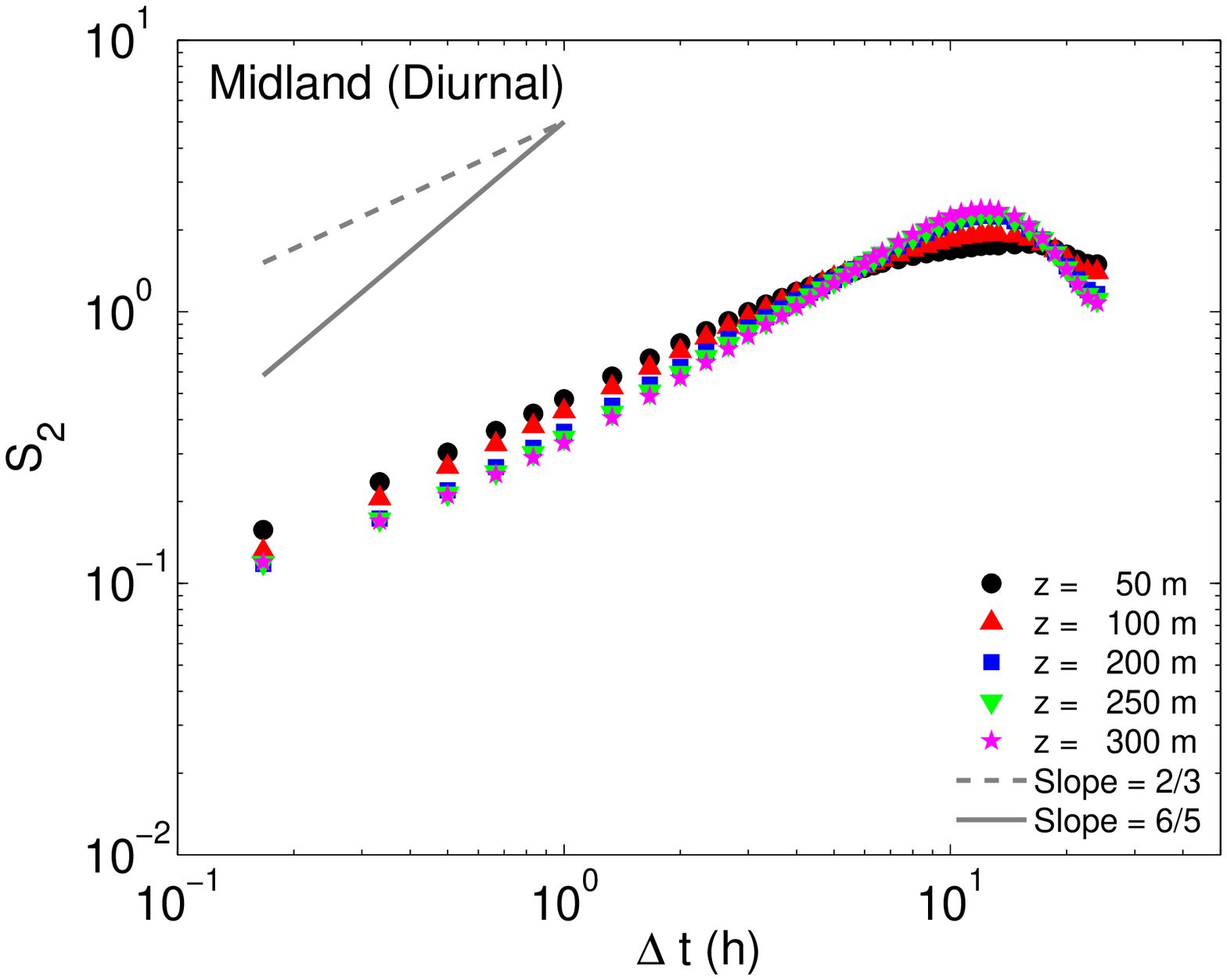}
\includegraphics[width=2.3in]{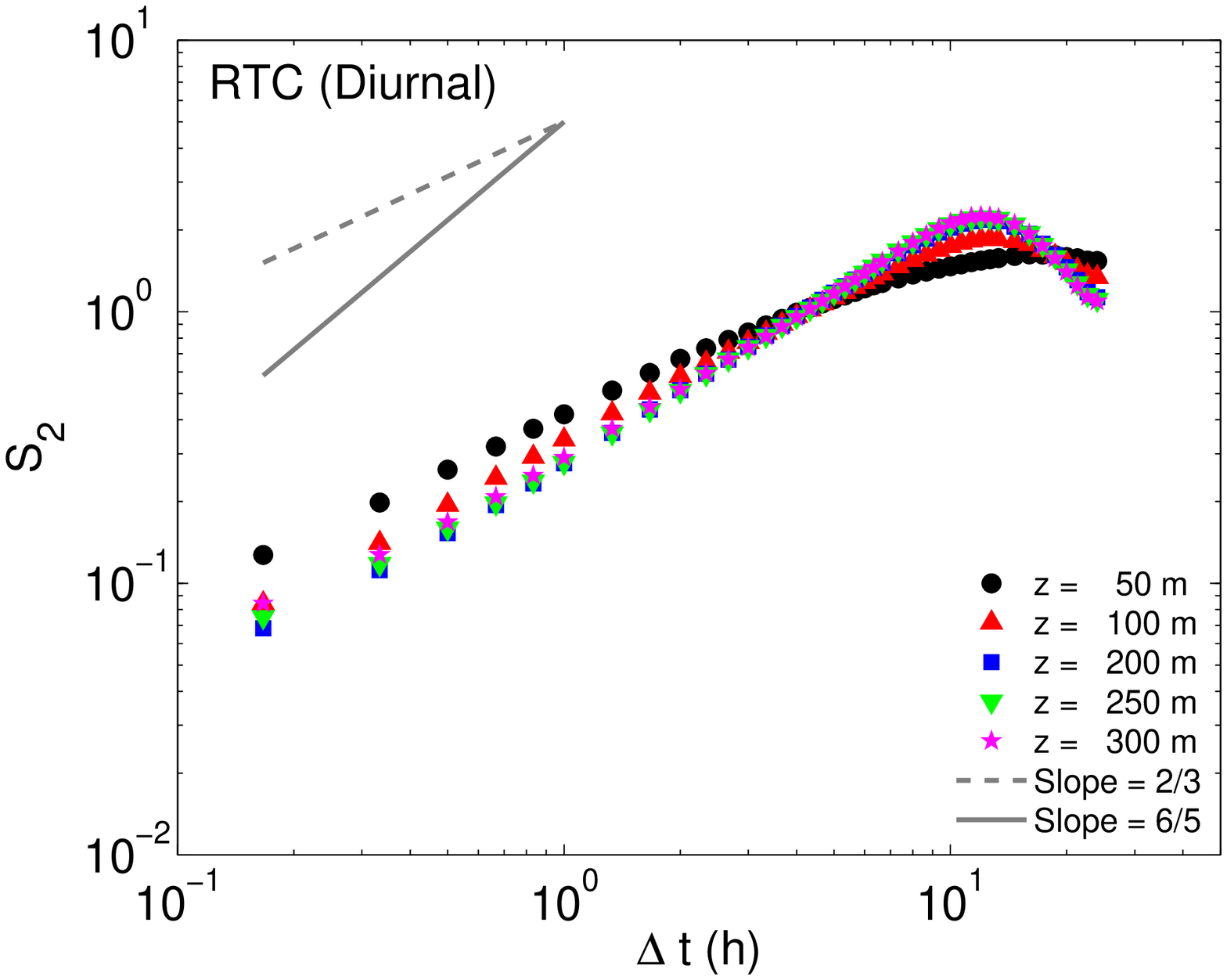}
\includegraphics[width=2.3in]{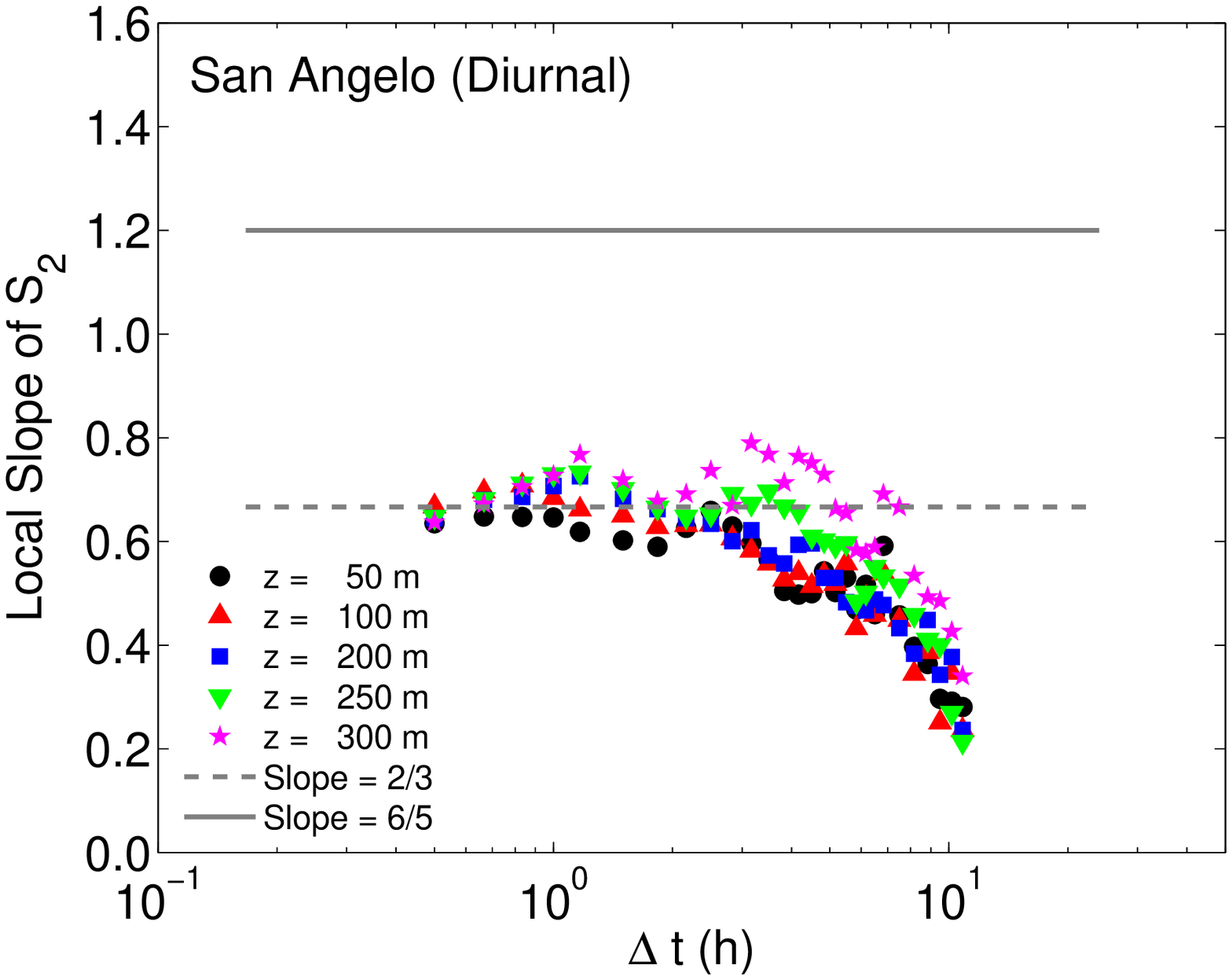}
\includegraphics[width=2.3in]{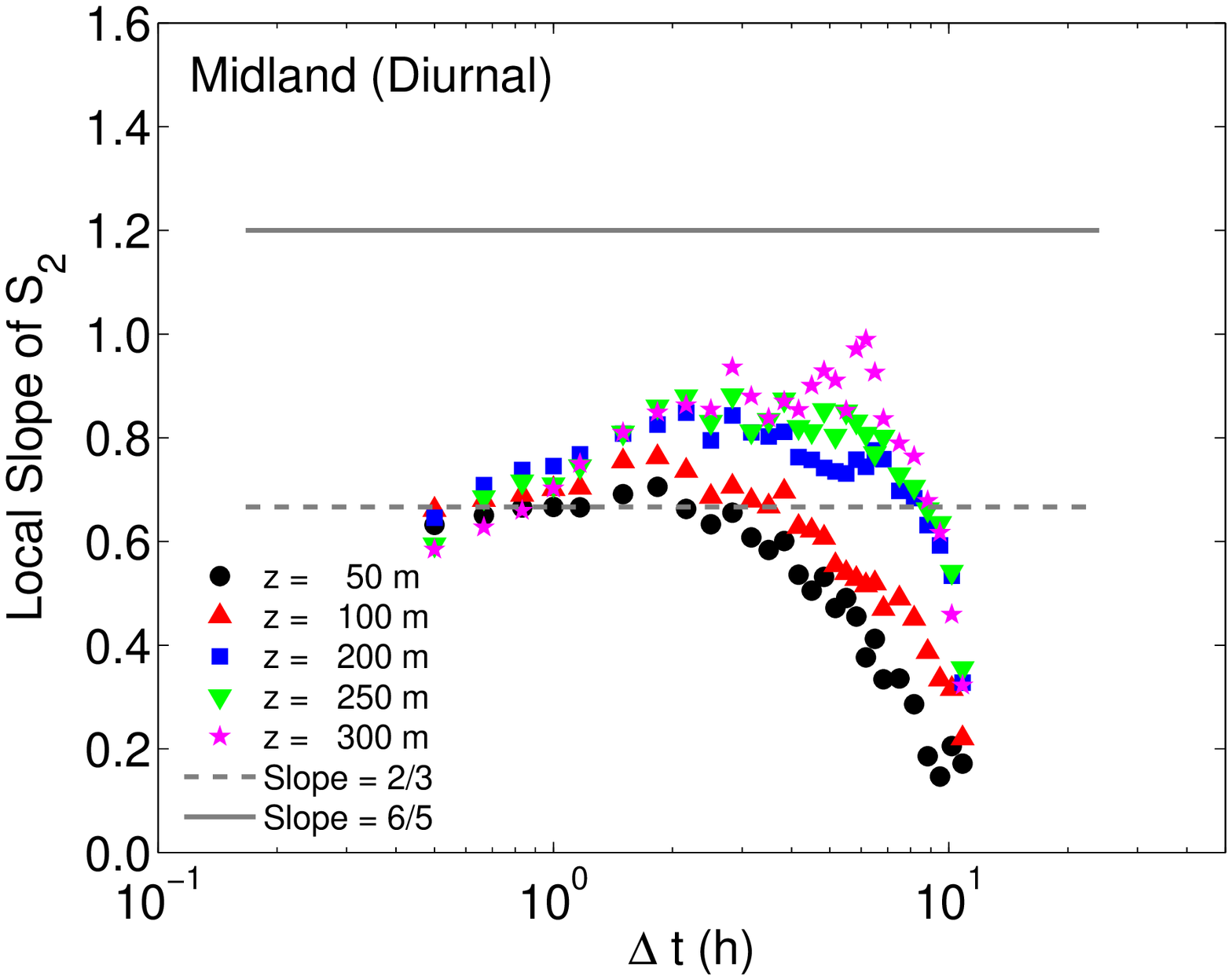}
\includegraphics[width=2.3in]{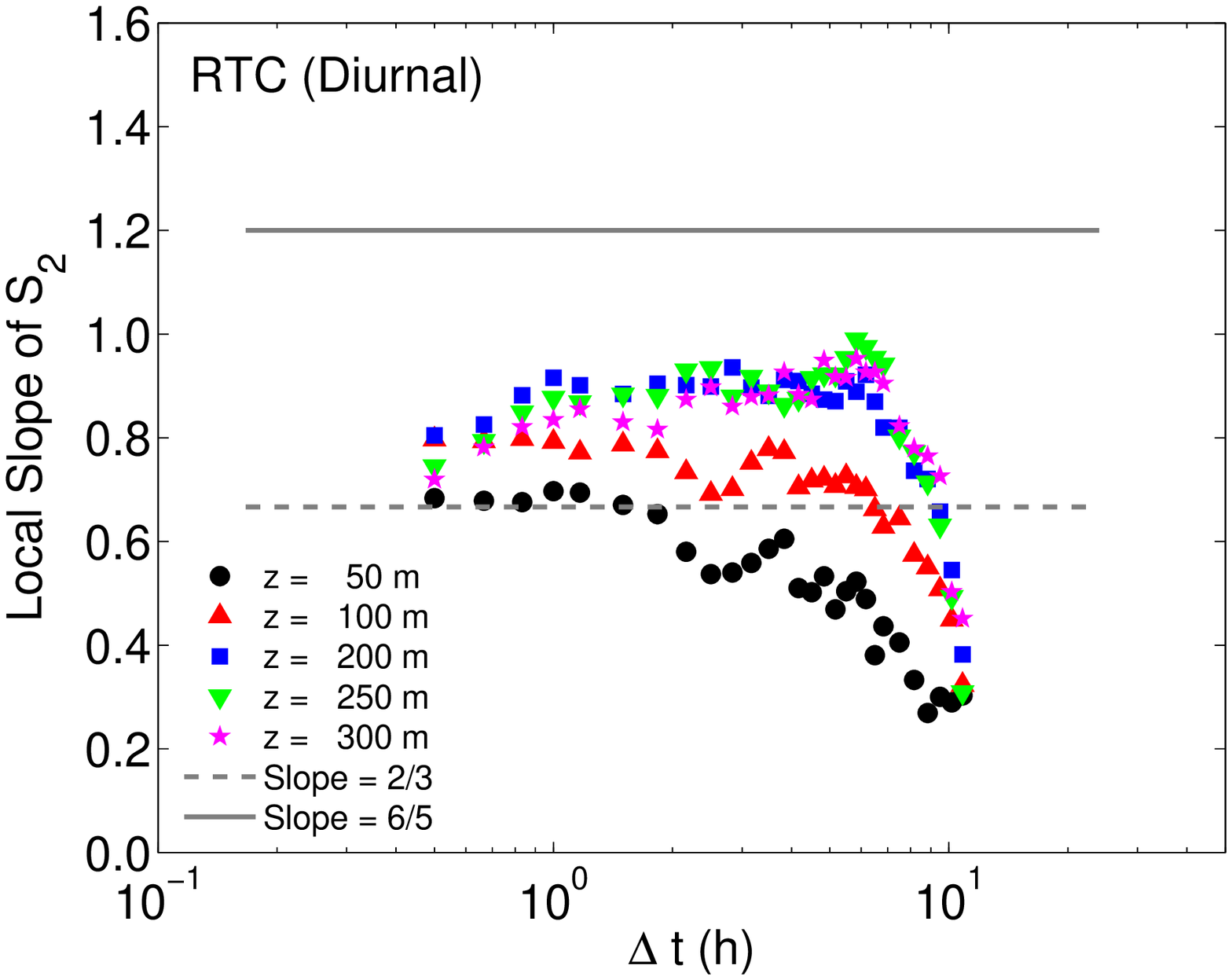}
\caption{\label{SF2-WTM} (Color Online) The second-order structure functions (top panel) and their corresponding local slopes (bottom panel) based on wind speed observations from the San Angelo (left panels), Midland (middle panels), and RTC (right panels) sites. The dashed and solid lines represent slopes corresponding to Kolmogorov's and Bolgiano's hypotheses, respectively.}
\end{figure*}

In Figs.~\ref{SF2-FINO1}--\ref{SF2-WTM}, the second-order structure functions ($S_2$) and their corresponding local slopes ($\zeta_2$, \footnote{A second-order central difference scheme (with non-uniform spacing) is used for the slope calculations.}) are shown for FINO 1, Cabauw, and WTM sodars. The following assertions can be made based on these figures: 

\begin{figure*}[!t]
\includegraphics[width=2.3in]{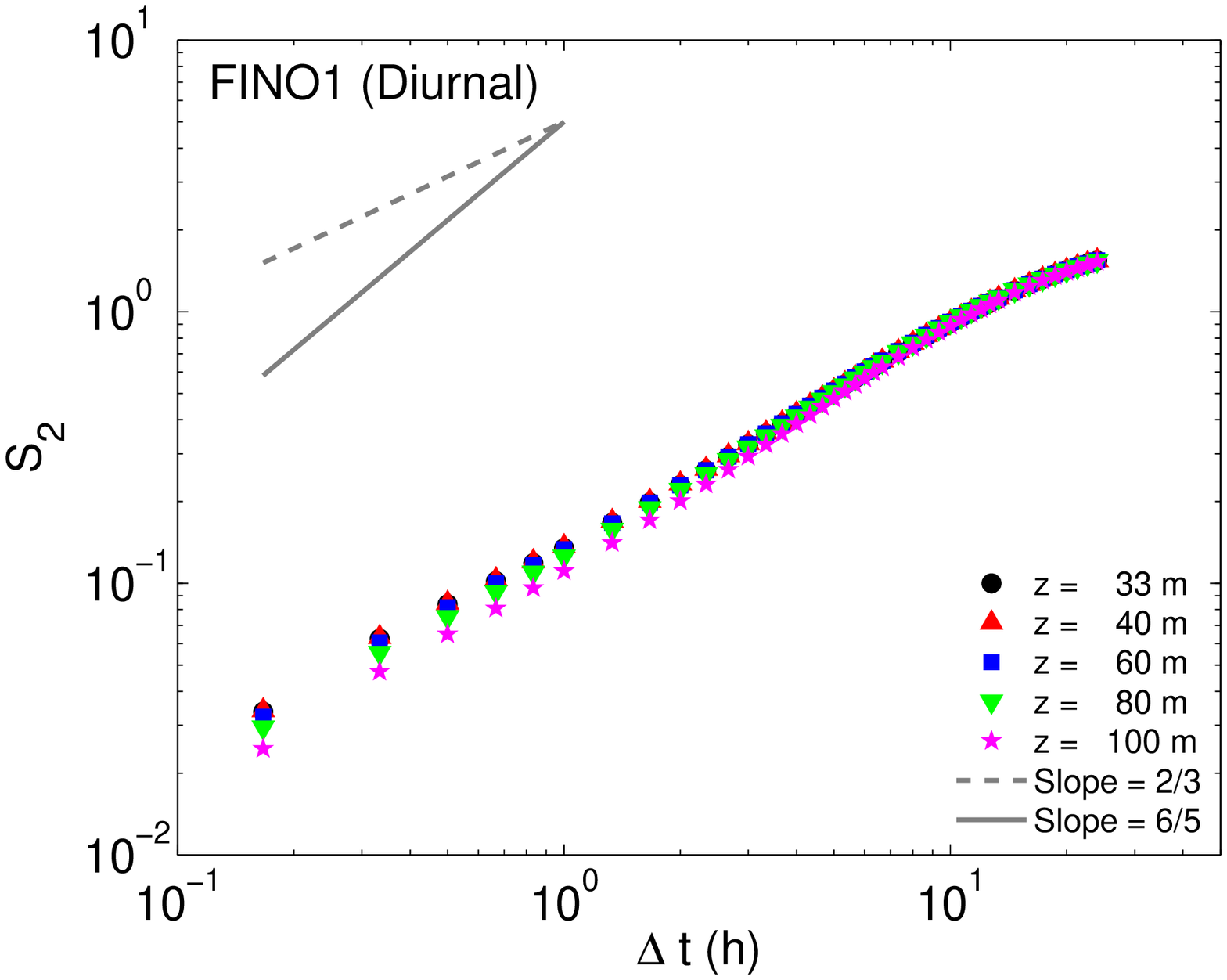}
\includegraphics[width=2.3in]{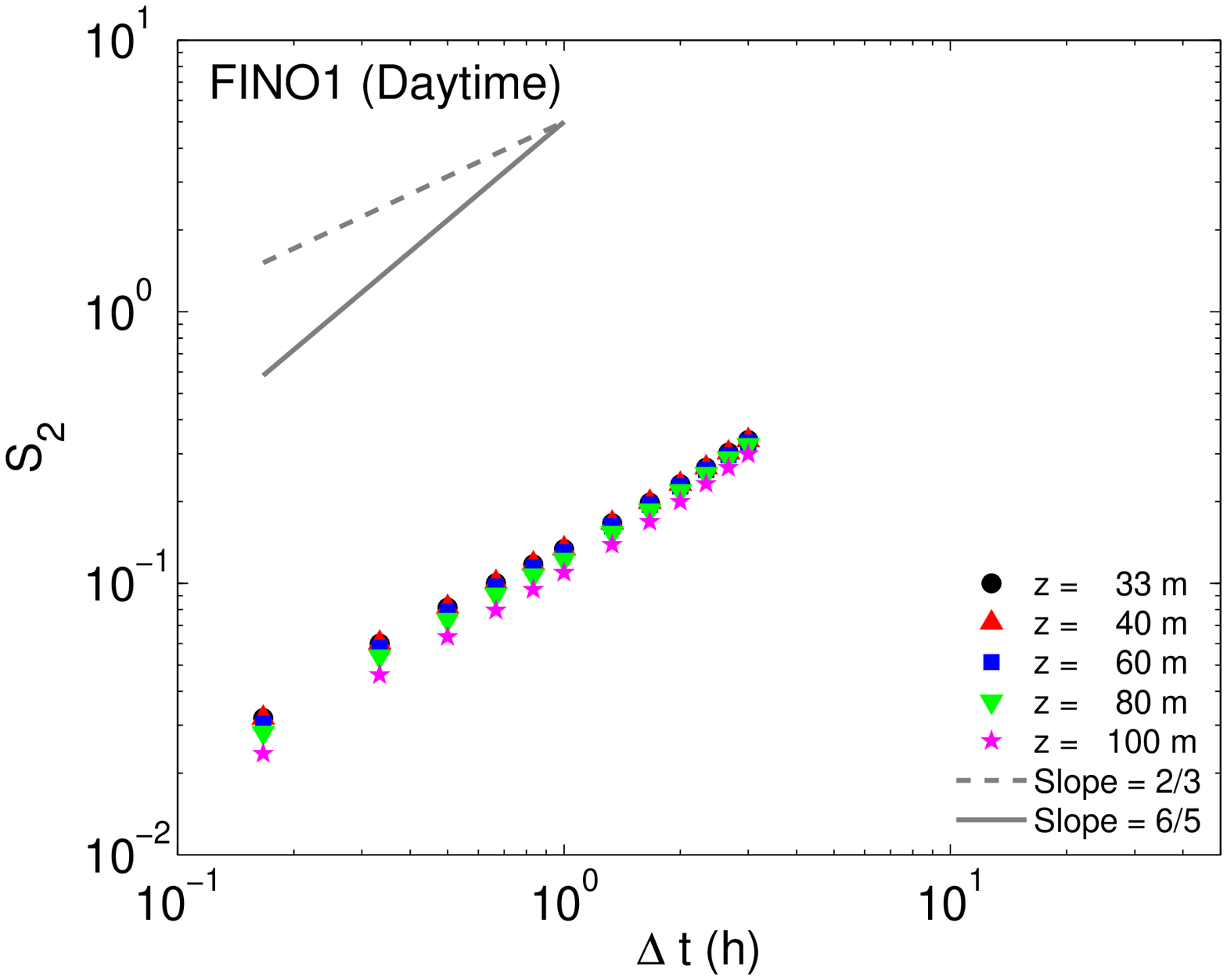}
\includegraphics[width=2.3in]{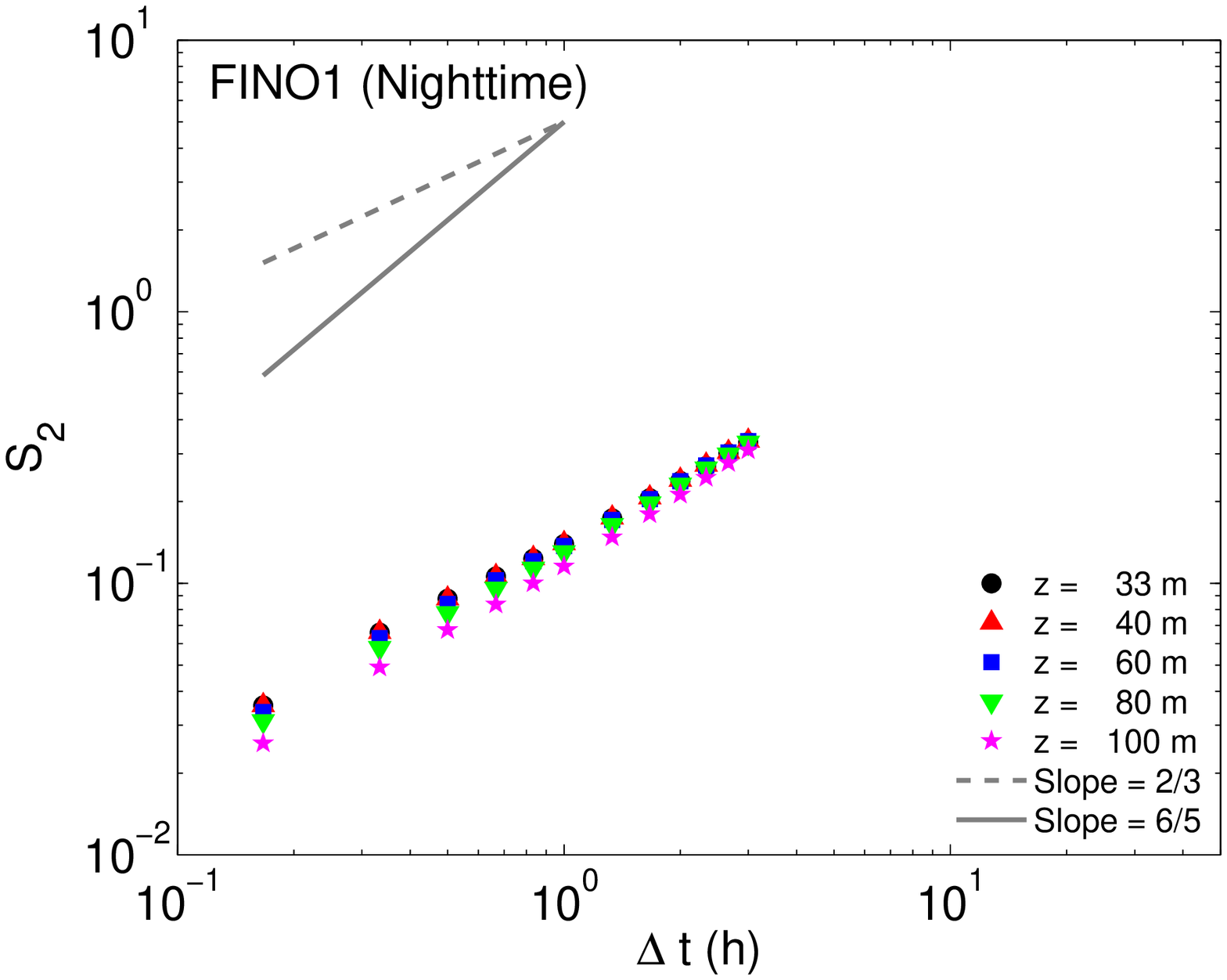}\\
\includegraphics[width=2.3in]{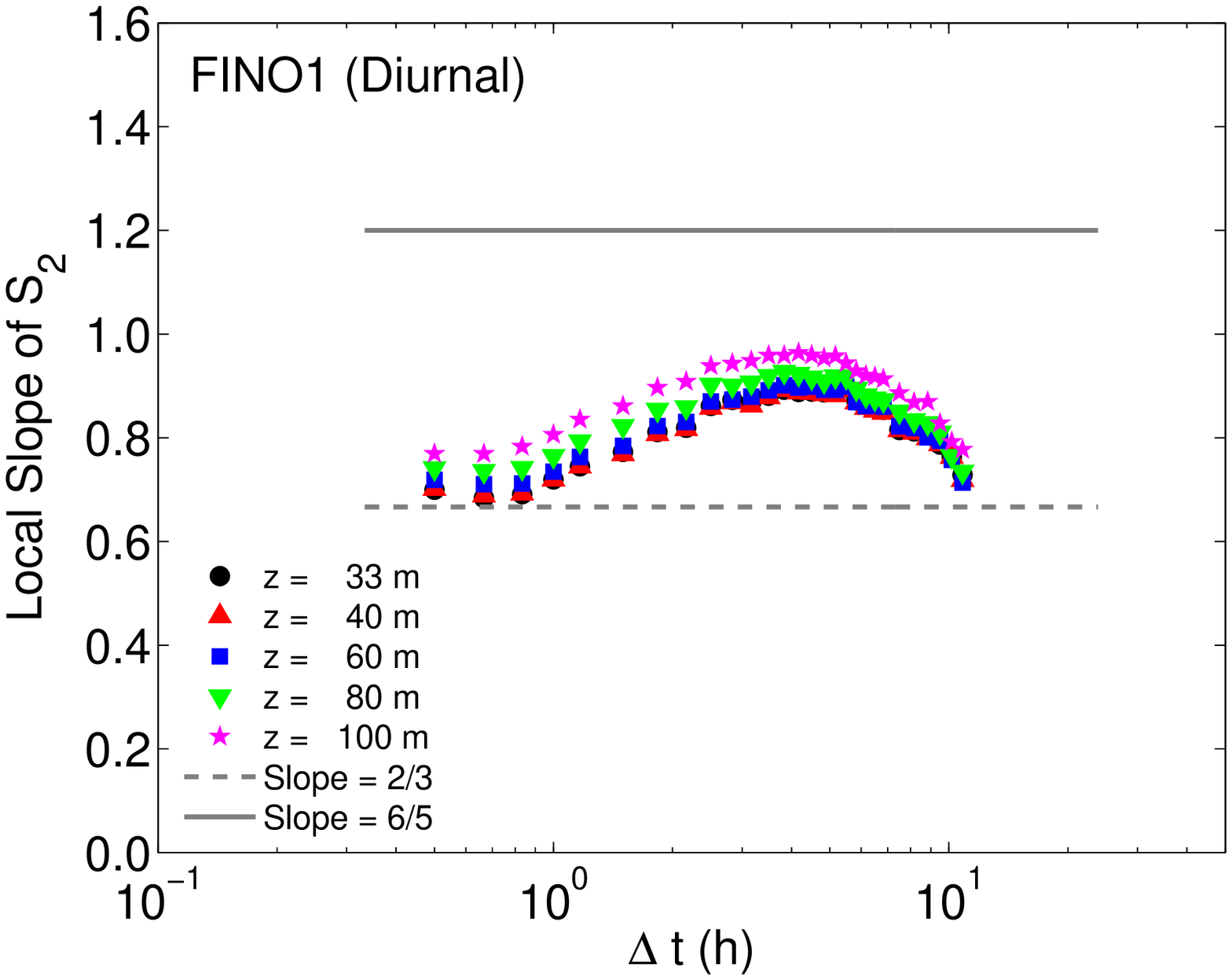}
\includegraphics[width=2.3in]{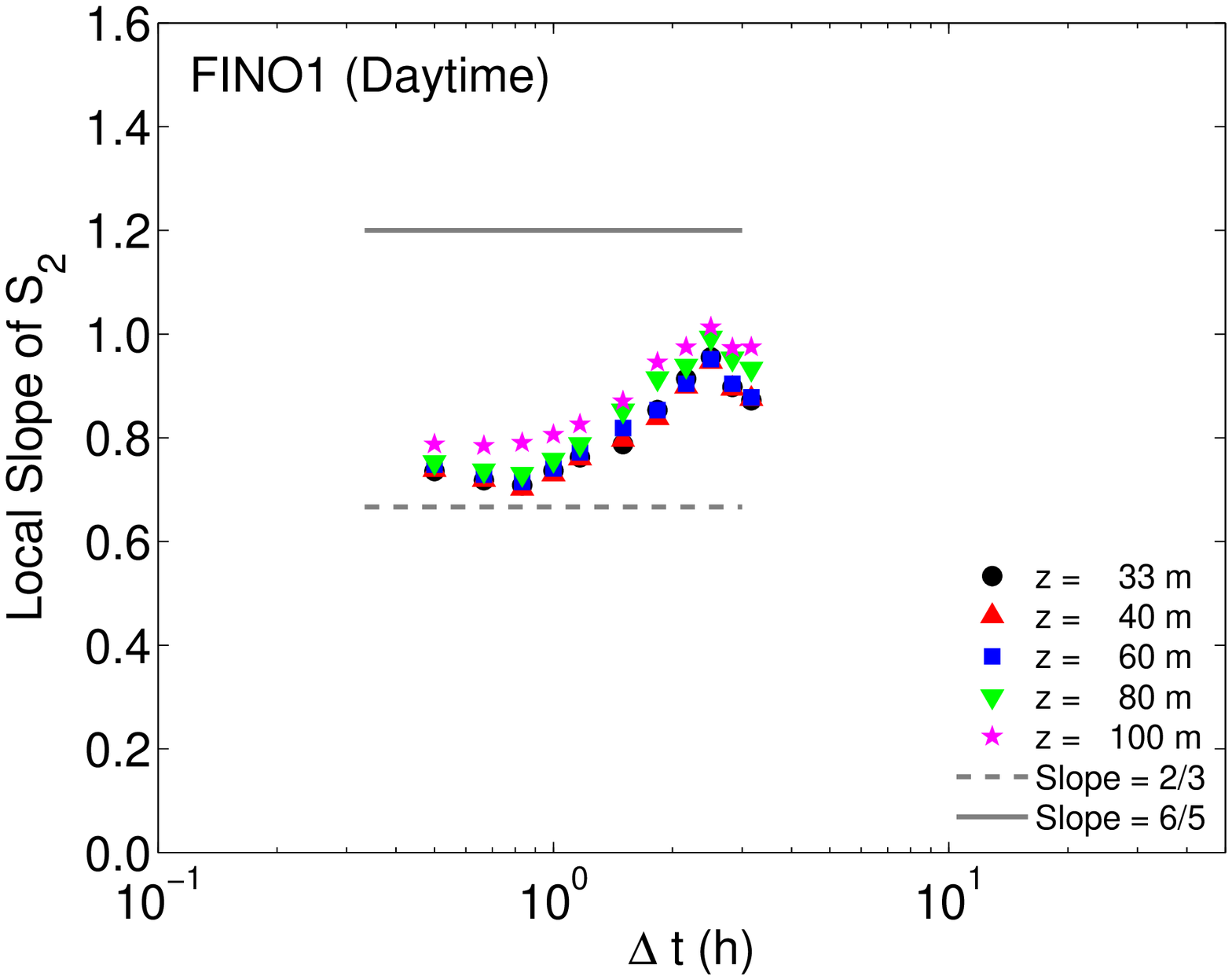}
\includegraphics[width=2.3in]{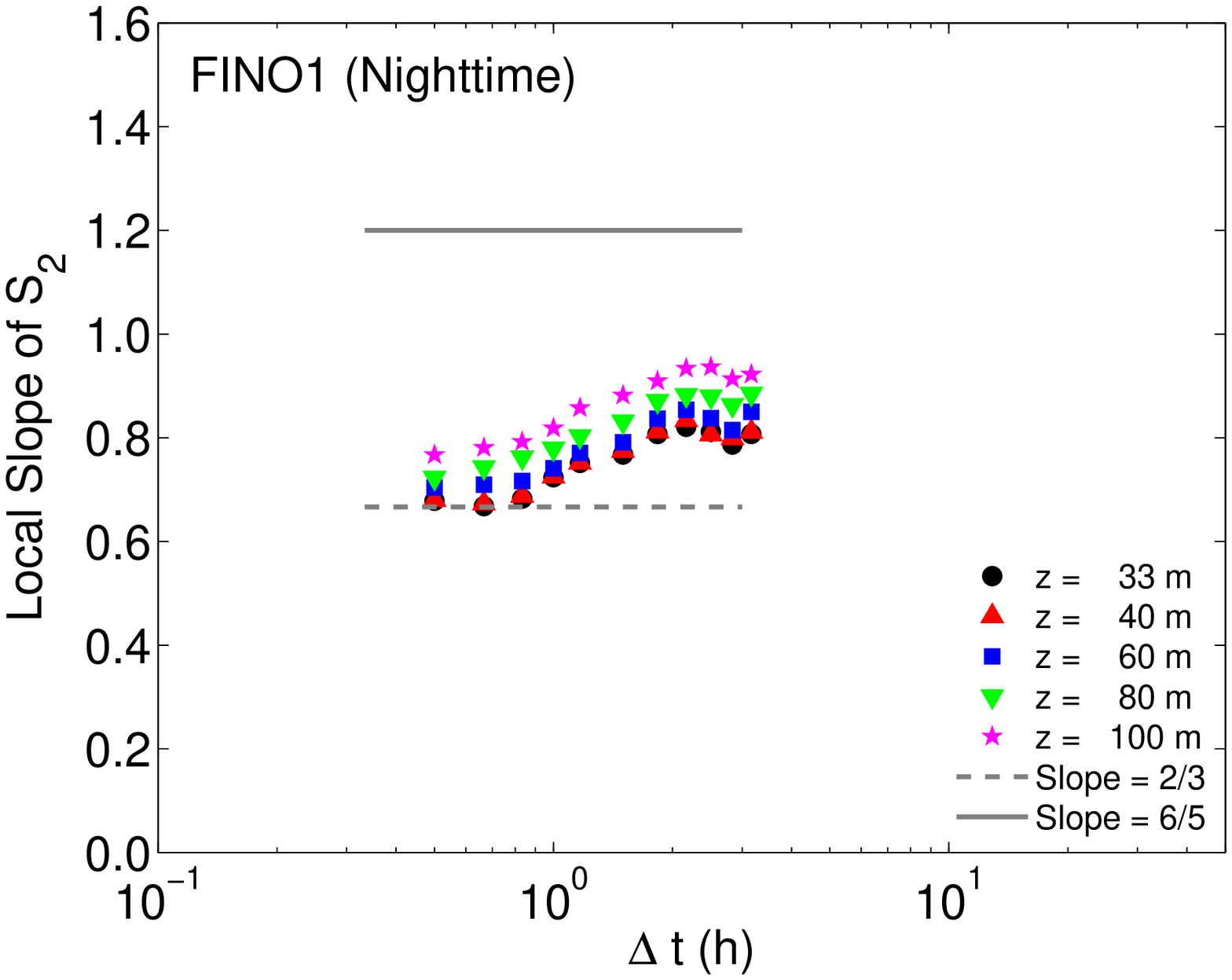}
\caption{\label{SF2-FINO1} (Color Online) The second-order structure functions (top panel) and their corresponding local slopes (bottom panel) based on wind speed observations from the FINO1 tower. The left, middle, and right panels represent diurnal, daytime (9--15 UTC), and nighttime (21--3 UTC) data, respectively. The dashed and solid lines represent slopes corresponding to Kolmogorov's and Bolgiano's hypotheses, respectively.}
\end{figure*}

\paragraph{FINO 1}

\begin{itemize}
\item Extended scaling regimes are clearly discernible in all the structure function plots. 
\item The $\zeta_2$ values increase monotonically with height. 
\item These slopes are lower for the smaller time-increments (i.e., $\Delta t < 1$ h) in comparison to the larger ones (i.e., 2 h $ < \Delta t < 6 $ h).
\item For lower heights and smaller time-increments, $\zeta_2$ values are close to the K-41 prediction (i.e., $\zeta_2 = 2/3$). Interestingly, the slopes remain significantly smaller than Bolgiano's prediction of 6/5 for all heights and for all time-increments.   
\item The scaling characteristics is (almost) indistinguishable between the daytime and nighttime cases. This behavior is somewhat expected at an offshore site where sea surface temperature (generally) exhibits a weak diurnal cycle. 
\end{itemize}

\paragraph{Cabauw}
\begin{itemize}
\item Once again, extended scaling regimes are clearly present in all the $S_2$ plots.
\item The $\zeta_2$ values increase with height with the exception of the lowest two tower levels. This discrepancy near the surface is possibly due to the disturbances caused by wind breaks and other sources of heterogeneity.
\item In contrast to the FINO 1 results, the scaling characteristics at Cabauw is strongly dependent on the time of the day. During nighttime (typically stably stratified condition over land), $\zeta_2$ values are much larger than their corresponding daytime (typically convective condition over land) values.   
\item The $\zeta_2$ values (approximately) equal to 2/3 for specific time-increment ranges and certain heights. However, these $\zeta_2$ values always remain much smaller than 6/5. 
\end{itemize}

\paragraph{WTM Sodars}
\begin{itemize}
\item Despite the limited sample size, for all the three sodar-based wind speed datasets, extended scaling regimes are visible. 
\item As before, the height-dependency of $\zeta_2$ is clearly noticeable for all the cases. However, this dependency is more pronounced at RTC followed by Midland. In contrast, the trend is much weaker at San Angelo. 
\item In the case of San Angelo and Midland, the $\zeta_2$ values are close to the K-41 value of 2/3 for $\Delta t$ up to 1 h for all the levels. However, at RTC, only wind speed at the lowest level portrays similar behavior. 
\item At RTC, for $z \ge 200$ m, the slopes are remarkably stable (long plateau); the $\zeta_2$ values are close to 1. 
\item At Midland, a dual-scaling behavior is discernible for $z > 100$ m. For $\Delta t \le 1$ h, $\zeta_2$ is approximately equal to 2/3; the slopes increase for larger $\Delta t$. 
\end{itemize}

In summary, $\zeta_2$ in the mesoscale range systematically depends on height and time-increment range. In the following section, we provide a physical explanation for the height-dependency by further analyzing the WTM sodar datasets. 

\section{Low-Level Jets}

\begin{figure}[h]
\includegraphics[width=3in]{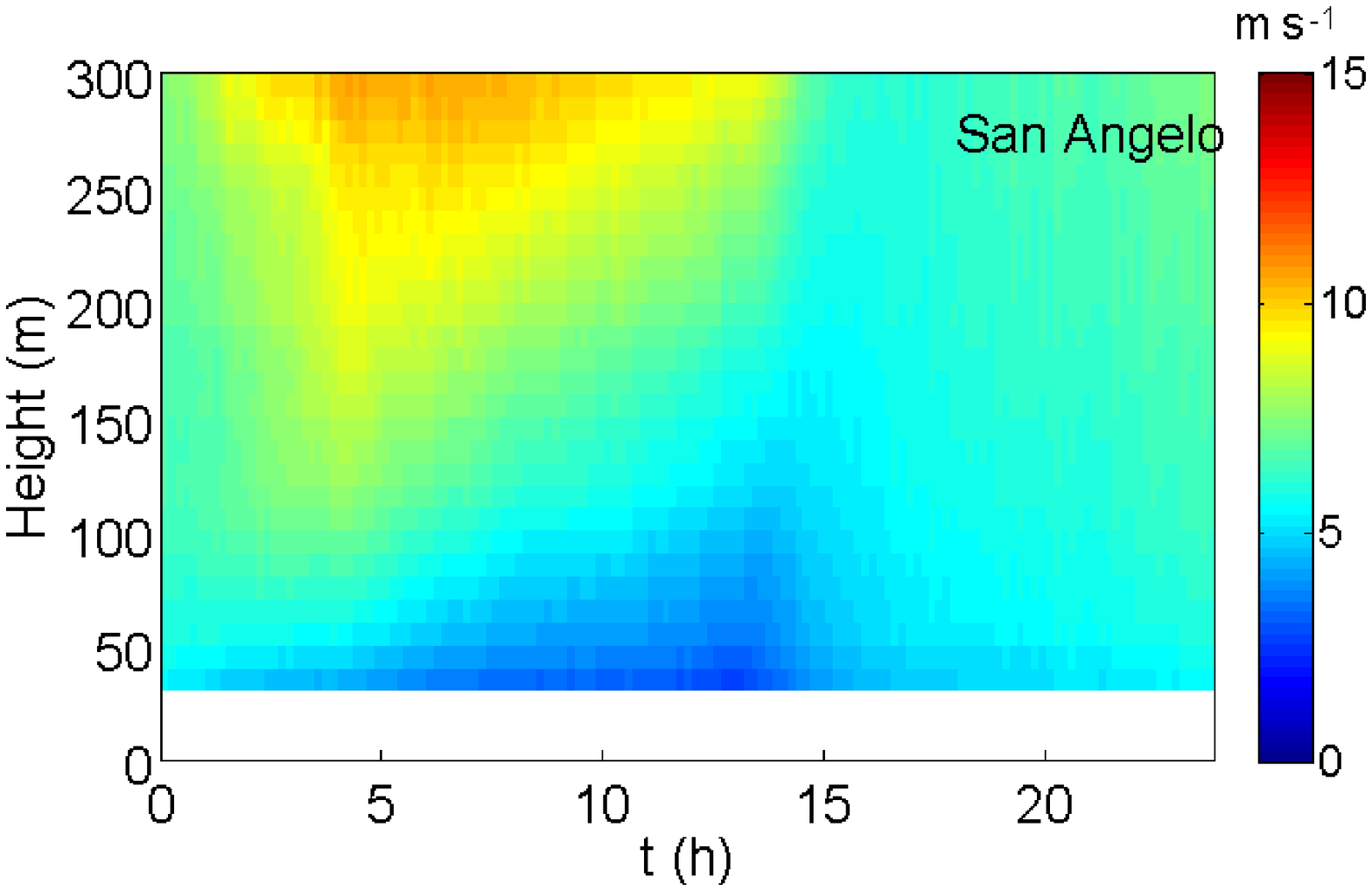}
\includegraphics[width=3in]{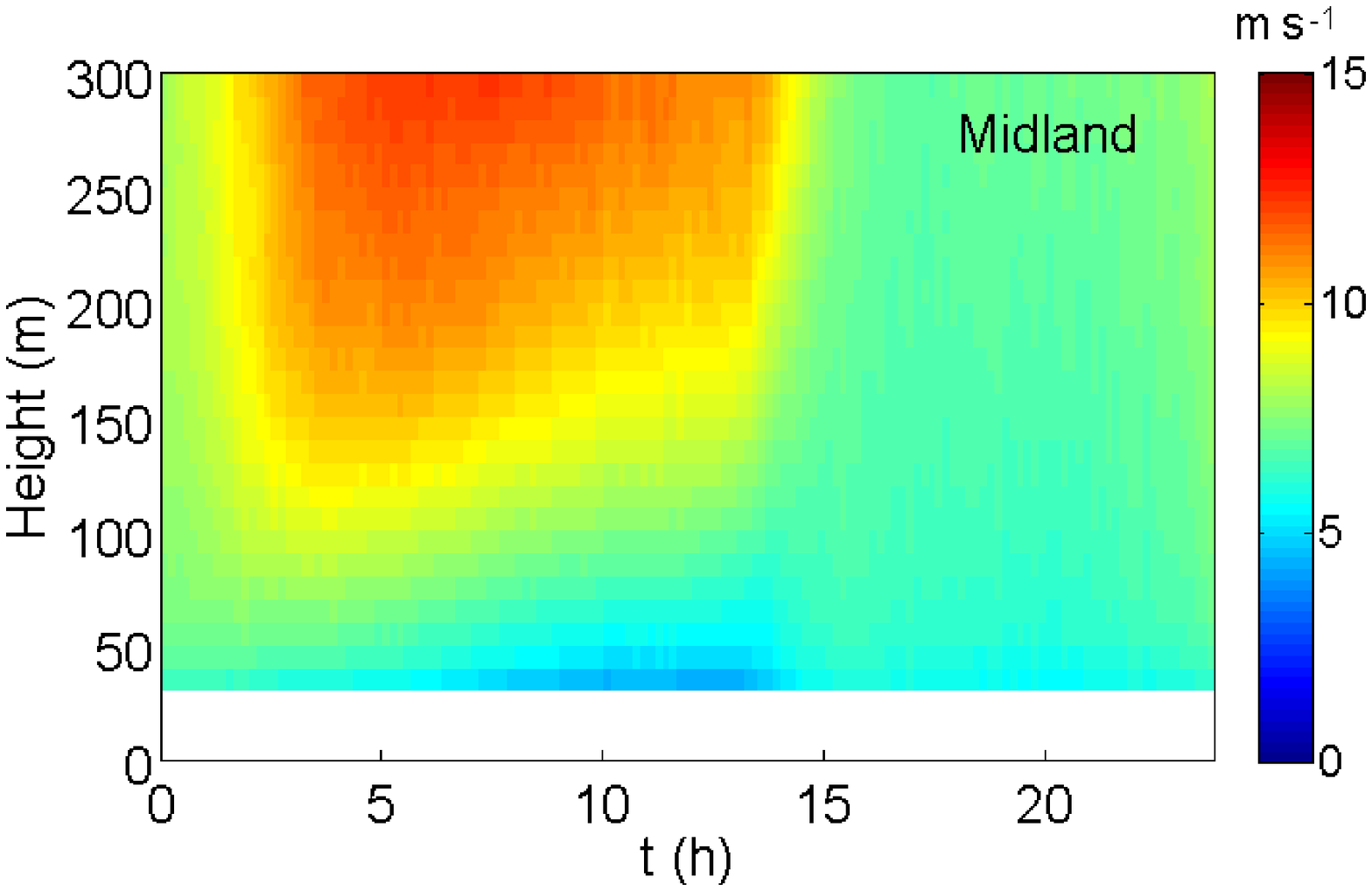}
\includegraphics[width=3in]{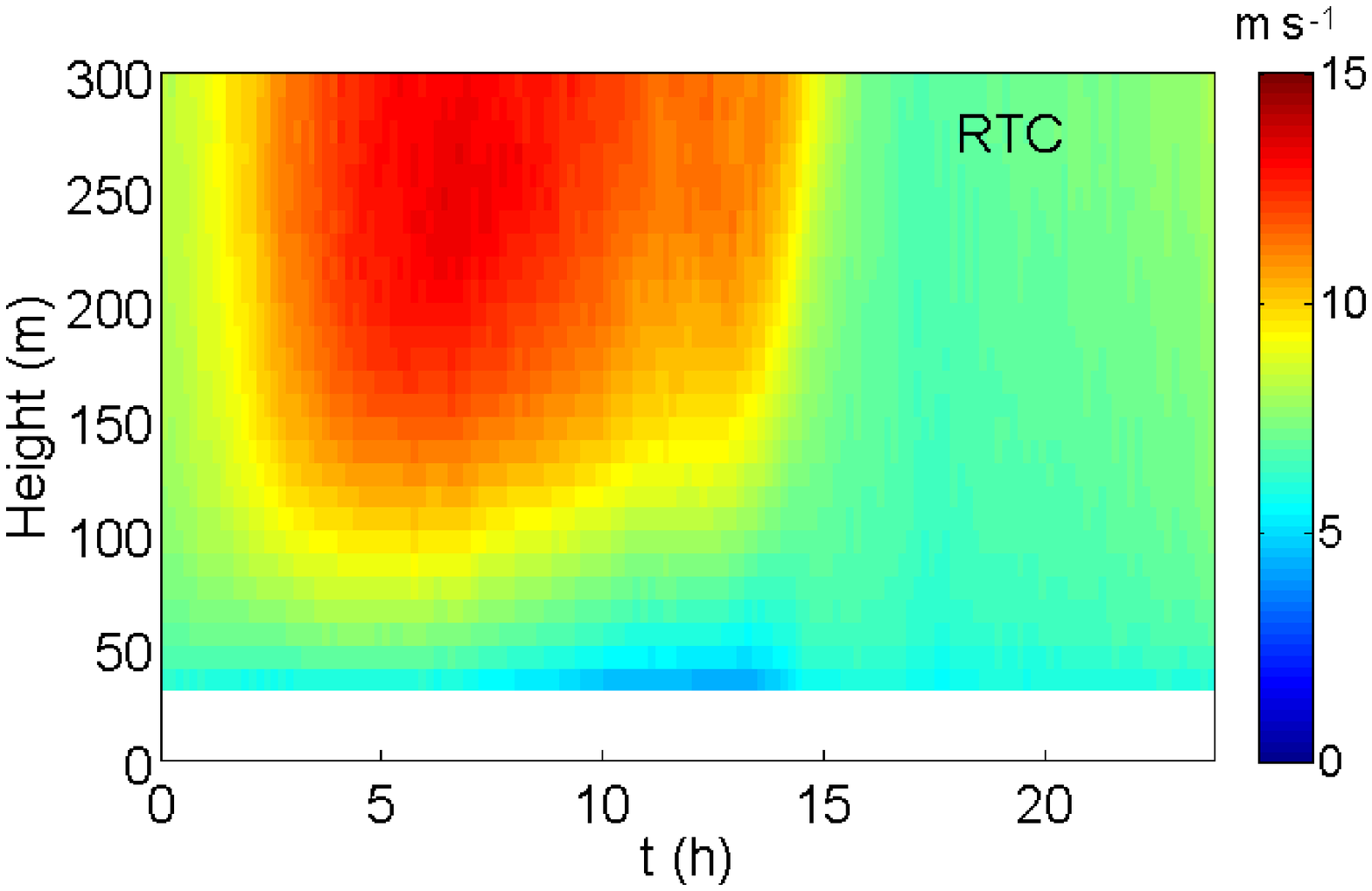}
\caption{\label{WSP} (Color Online) Time-height plots of averaged wind speed at the San Angelo (top panel), Midland (middle panel), and RTC (bottom panel) measurement sites. Time is in UTC. Height is above ground level.}
\end{figure}

The WTM sodars are located in the West Texas Panhandle region of the US, one of the largest semiarid regions in the world. This region is gently sloping, homogeneous, and sparsely vegetated. Nocturnal low-level jets (LLJs; \cite{Blackadar1957,bonn68,sist78,Songetal2005,rife10}) occur frequently over this region. These LLJs and associated wind speed maxima are one of the most important reasons for the abundance of wind farms, as well as significant nighttime wind power production, over this region \cite{storm2009evaluation,stor10,wilczak2014wind}.

In order to explain the height-dependency of $\zeta_2$, time-height plots of averaged (over an eight month period) wind speed at the measurement sites are shown in Fig.~\ref{WSP}. During the daytime, the wind fields are well-mixed as would be physically expected and the differences across the sites are marginal. However, the nighttime scenarios are completely different. Even though the LLJs are frequently present at all of the sites, their locations are significantly different. It appears that the distance between the LLJ and the underlying surface decreases with increasing site elevation (see Table~\ref{tab:table1}). Similar observations were reported at a different site over the US Great Plains by Song et al. \cite{Songetal2005}. 

One of the primary mechanisms for the formation of the LLJs is the so-called inertial oscillation \cite{Blackadar1957,van2010conceptual,parish2010role}. According to this mechanism, the buoyancy effects largely contribute to the formation and dynamical evolution of the LLJs. Thus, based on Fig.~\ref{WSP} (please refer to  Appendix~\ref{stratification} for further evidence), one could conjecture that the buoyancy effects will be increasing from San Angelo to RTC (via Midland). 

The results in Fig.~\ref{SF2-WTM} which show the scaling exponents increasing from San Angelo to RTC (via Midland) are definitely in line with this conjecture. In other words, the height-dependency of the scaling exponents is likely due to buoyancy effects. Similar height-dependency was reported in a wind-tunnel study  \cite{ruiz2000scaling}. In that case, enhanced shear near the surface decreased the scaling exponents from the corresponding inertial-range values. In our case, the trend is reversed due to the buoyancy effects. 

\section{Extended Self-Similarity}

\begin{figure*}[!ht]
\includegraphics[width=2.3in]{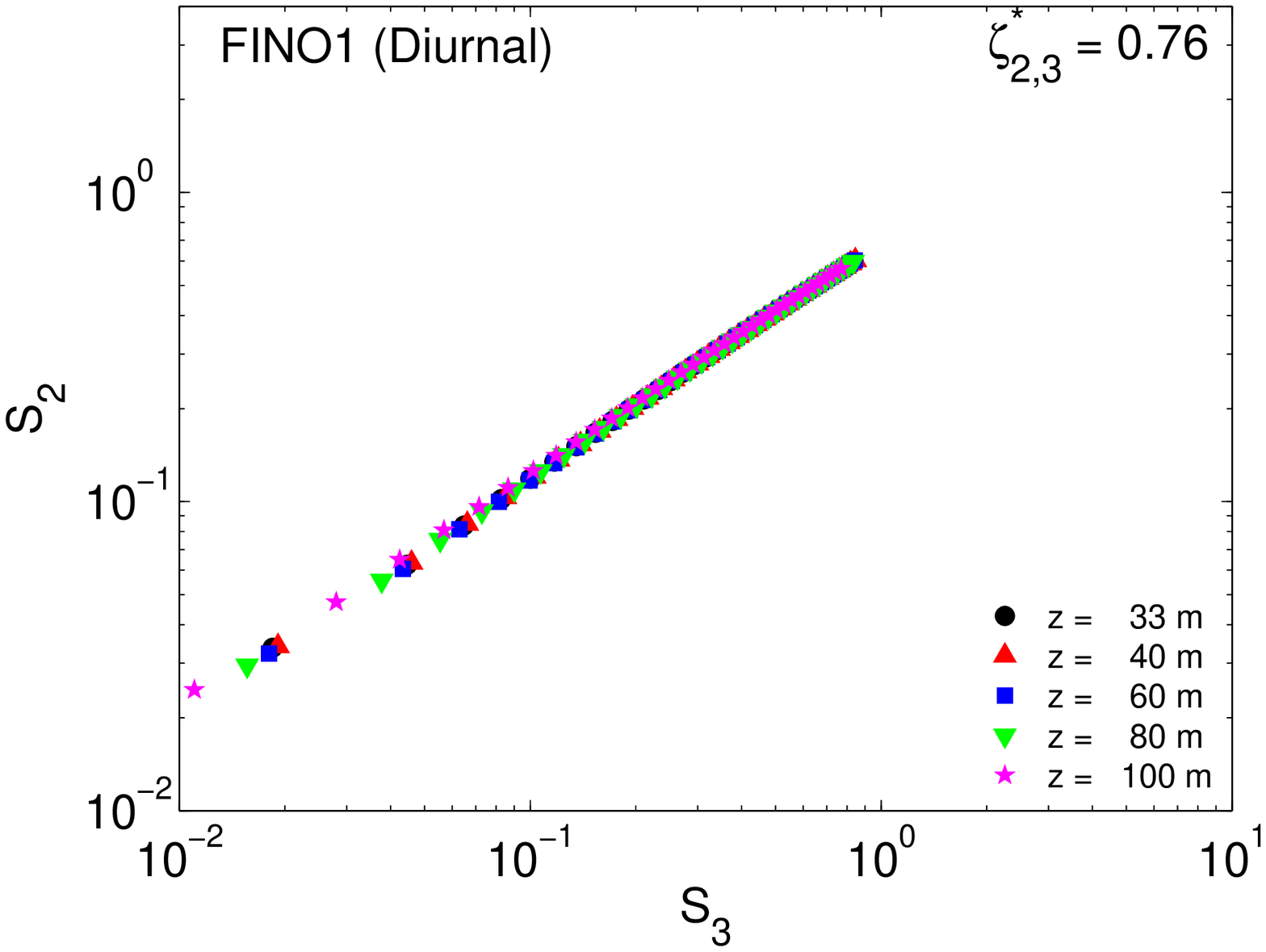}
\includegraphics[width=2.3in]{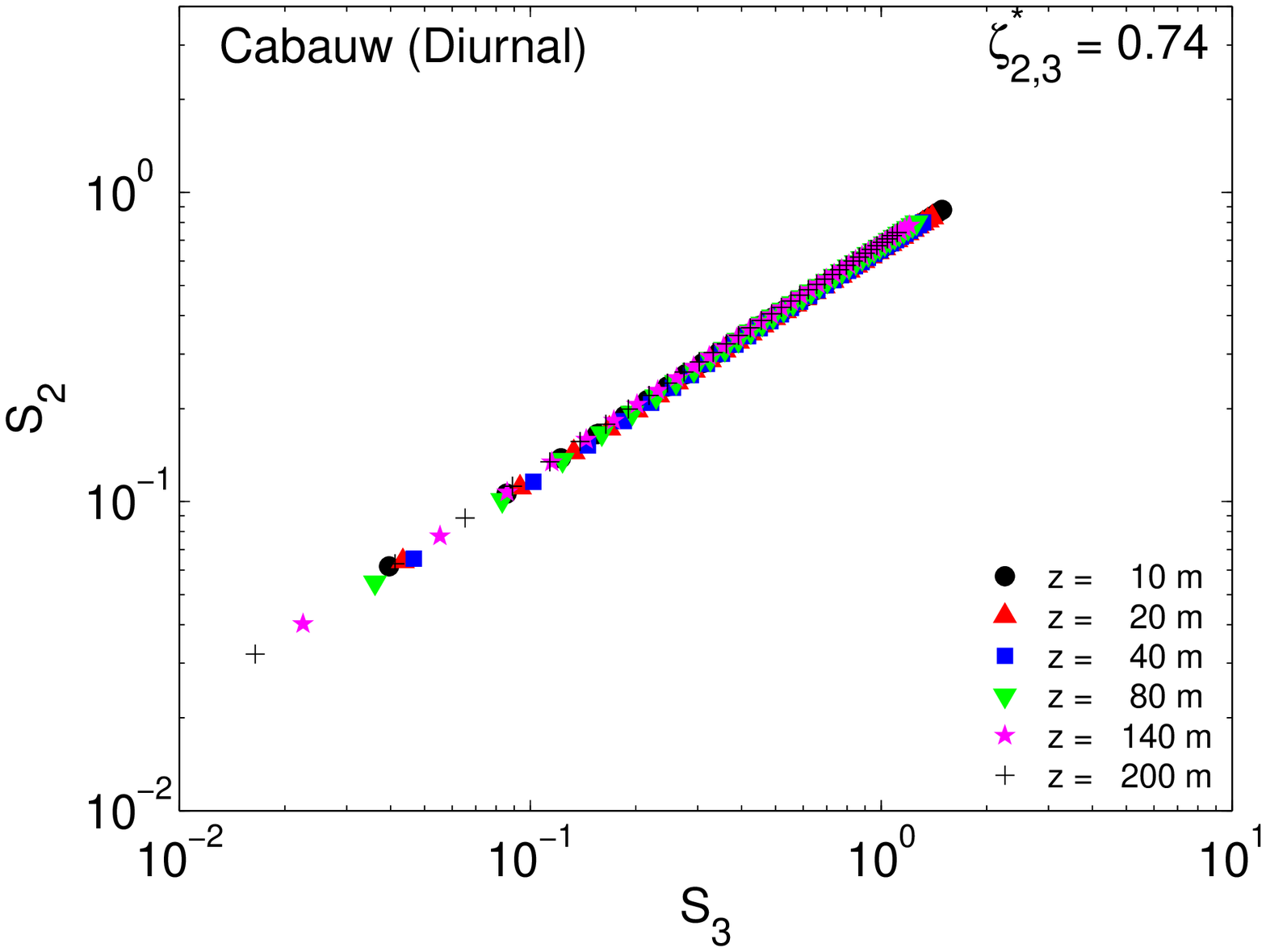}
\includegraphics[width=2.3in]{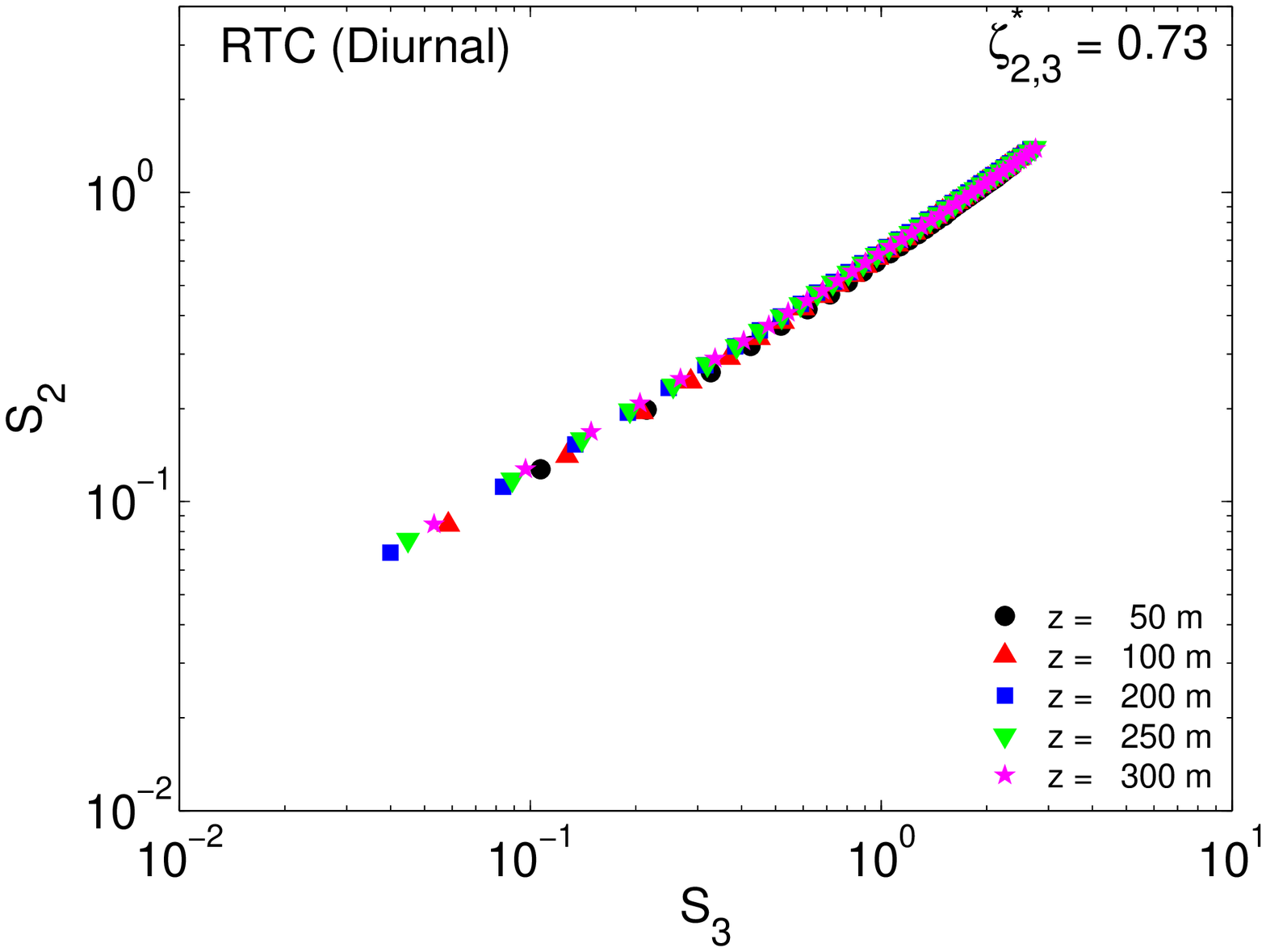}\\%
\includegraphics[width=2.3in]{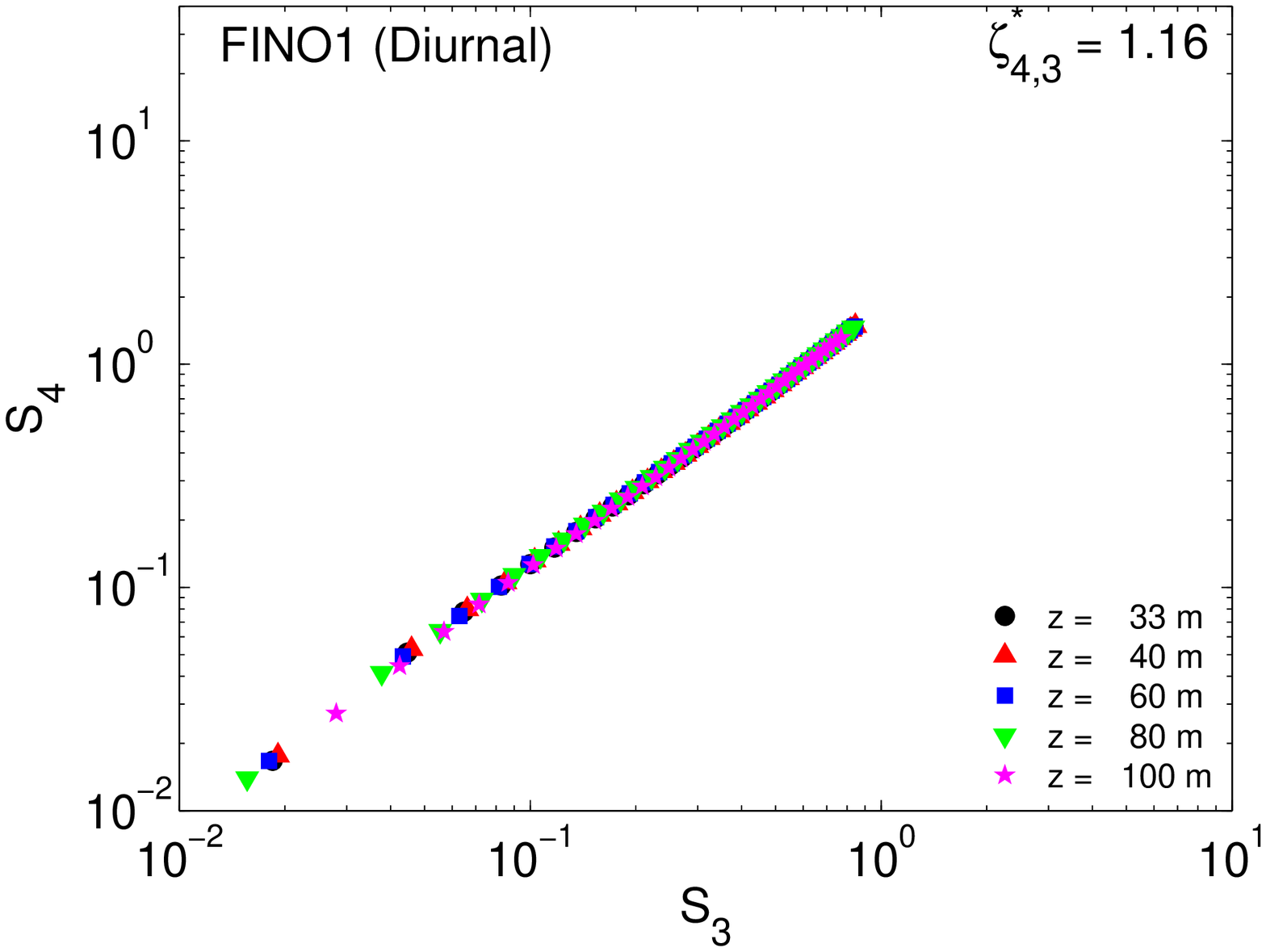}
\includegraphics[width=2.3in]{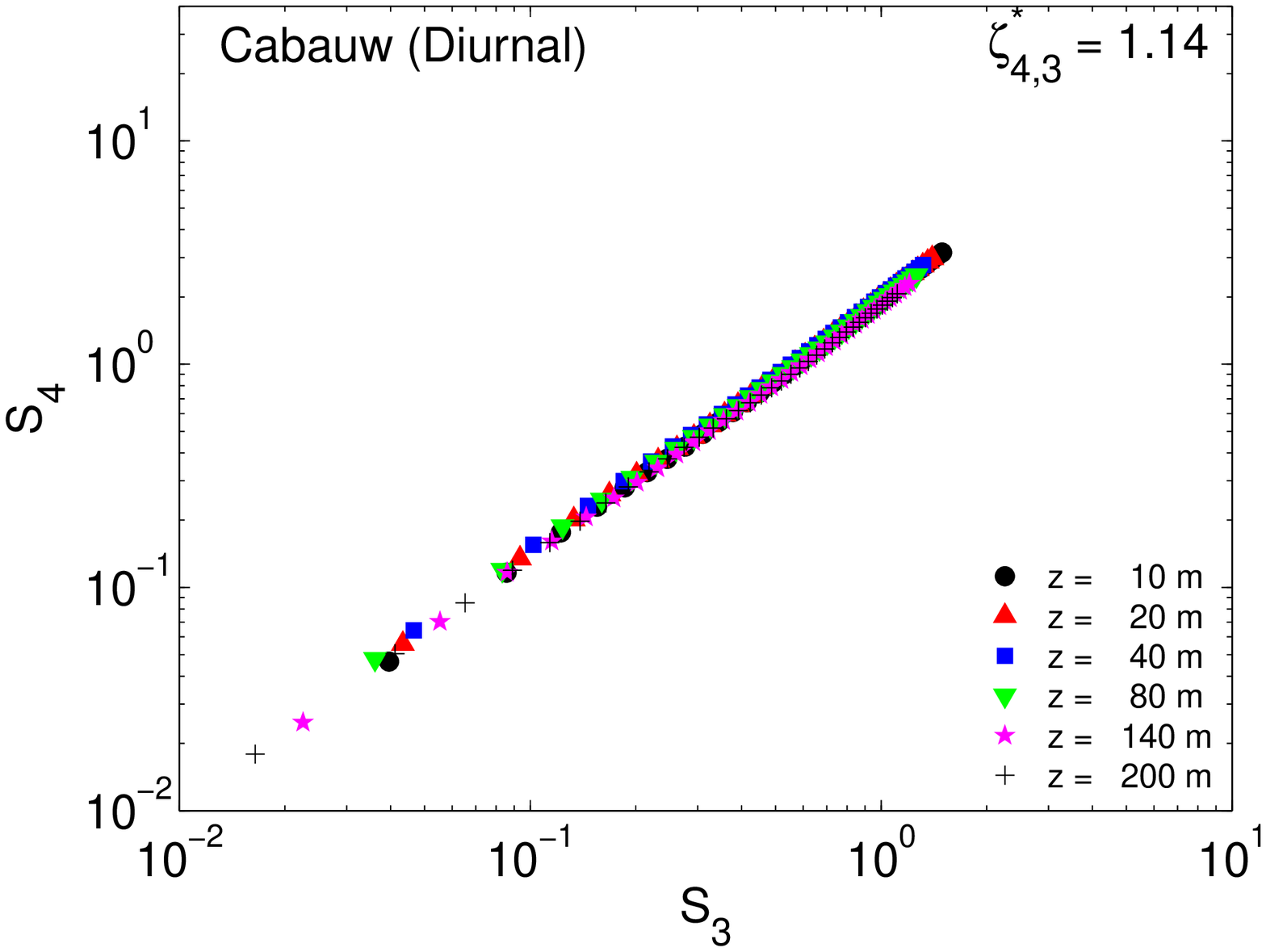}
\includegraphics[width=2.3in]{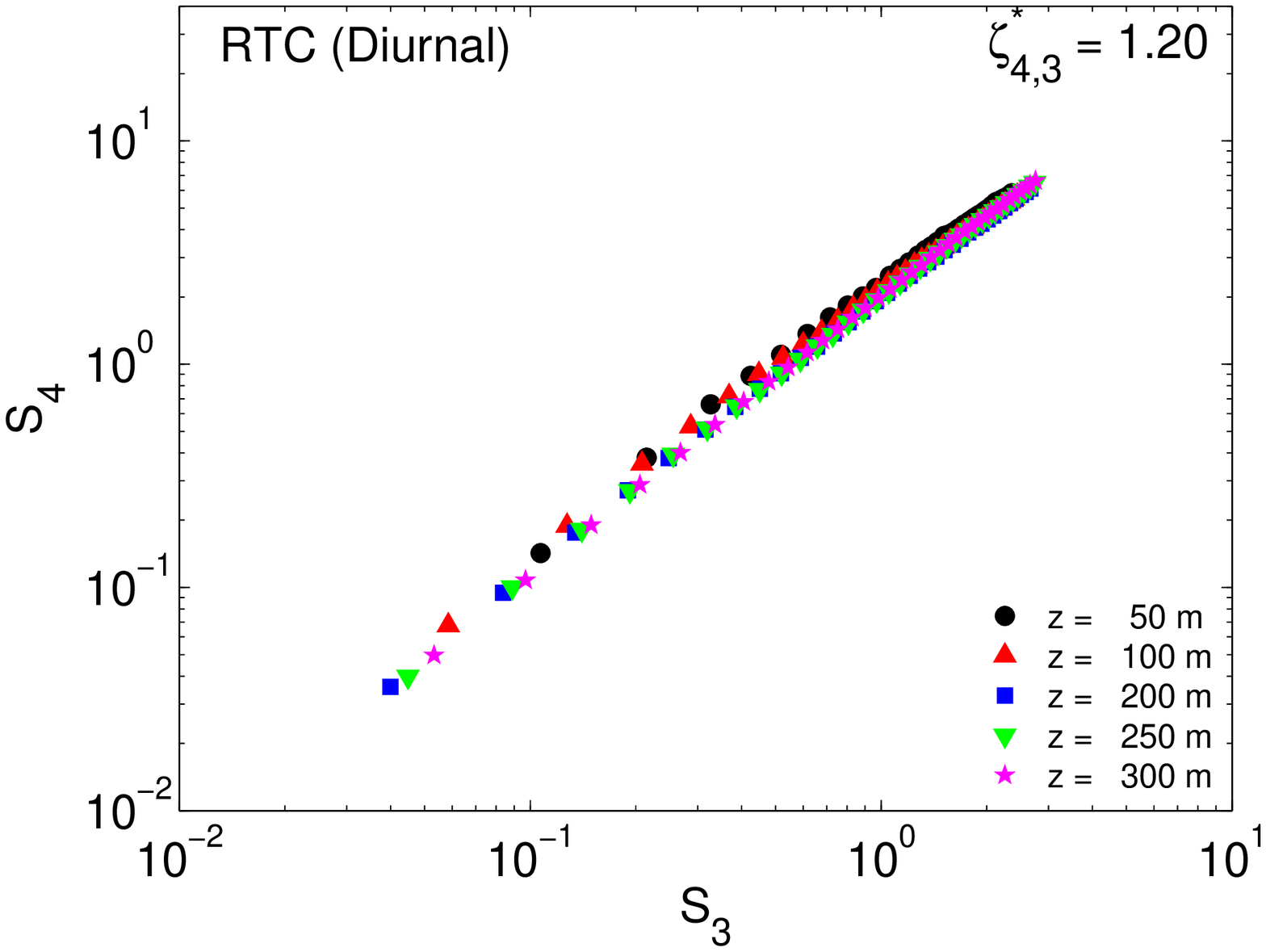}
\caption{\label{ESS} (Color Online) Top-panel: the variation of the second-order structure function with respect to the third-order structure function. Bottom-panel: the variation of the fourth-order structure function with respect to the third-order structure function. The left, middle, and right panels represent FINO 1, Cabauw, and RTC cases, respectively. The overall relative exponents $\zeta^{*}_{p,q}$ are reported on the top right corner of each plot.}
\end{figure*}

To further characterize scaling in the mesoscale regime, we next invoke the extended self-similarity (ESS) framework by Benzi and his co-workers \cite{benzi1993extended}. Numerous studies (see \cite{arneodo1996structure} and the references therein) have demonstrated the strength of ESS in terms of identifying scaling regimes even when the traditional structure function approach completely fails. In Fig.~\ref{ESS}, we show $S_2$ vs. $S_3$ and $S_4$ vs. $S_3$. All the structure function values corresponding to $\Delta t \le$ 6 h are considered. Remarkably, the height-dependency and dual-scaling behaviors have (almost) disappeared in these plots. Similar masking effects of ESS were also reported by Ruiz-Chavarria et al.~\cite{ruiz2000scaling} and Aivalis et al.~\cite{aivalis2002temperature} for completely different types of flows. 

Based on the ESS plots, one can calculate the relative scaling exponents $\zeta^*_{p,q} = \zeta_p/\zeta_q$. These values are reported in Fig.~\ref{ESS}. Given their close agreement across diverse sites, we speculate that $\zeta^*_{p,q}$ values are probably quasi-universal. Please note that these values are marginally different (more intermittent) from the commonly reported inertial-range values in the turbulence literature \cite{frisch1995turbulence}: $\zeta^*_{2,3} = 0.70$ and $\zeta^*_{4,3} = 1.28$.  In other words, the scaling characteristics of the near-surface wind field in the mesoscale regime appears to be similar to the inertial-range, but not exactly the same. This finding is in line with the recent literature \cite{muzy2010intermittency,baile2010spatial,liu2013cascade}. 

\section{Concluding Remarks}

In this study, we provide empirical evidence of quasi-universal scaling of ABL wind speed in the mesoscale range. This quasi-universality is only evident when the ESS framework is employed. Without this framework, the scaling exponents portray systematic dependence on height and buoyancy effects. Further observational data analyses are needed in order to gain further confidence on these noteworthy findings. If the ESS-based quasi-universal scaling holds for other geographical and meteorological regions (e.g., polar region, complex terrain), then it can be utilized as a benchmark for the development of next-generation planetary boundary layer parameterizations.  

\appendix
\section{Effects of Diurnal Cycles on the Scaling Exponents}
\label{filtering}
\begin{figure*}[htpb]
\includegraphics[width=2.3in]{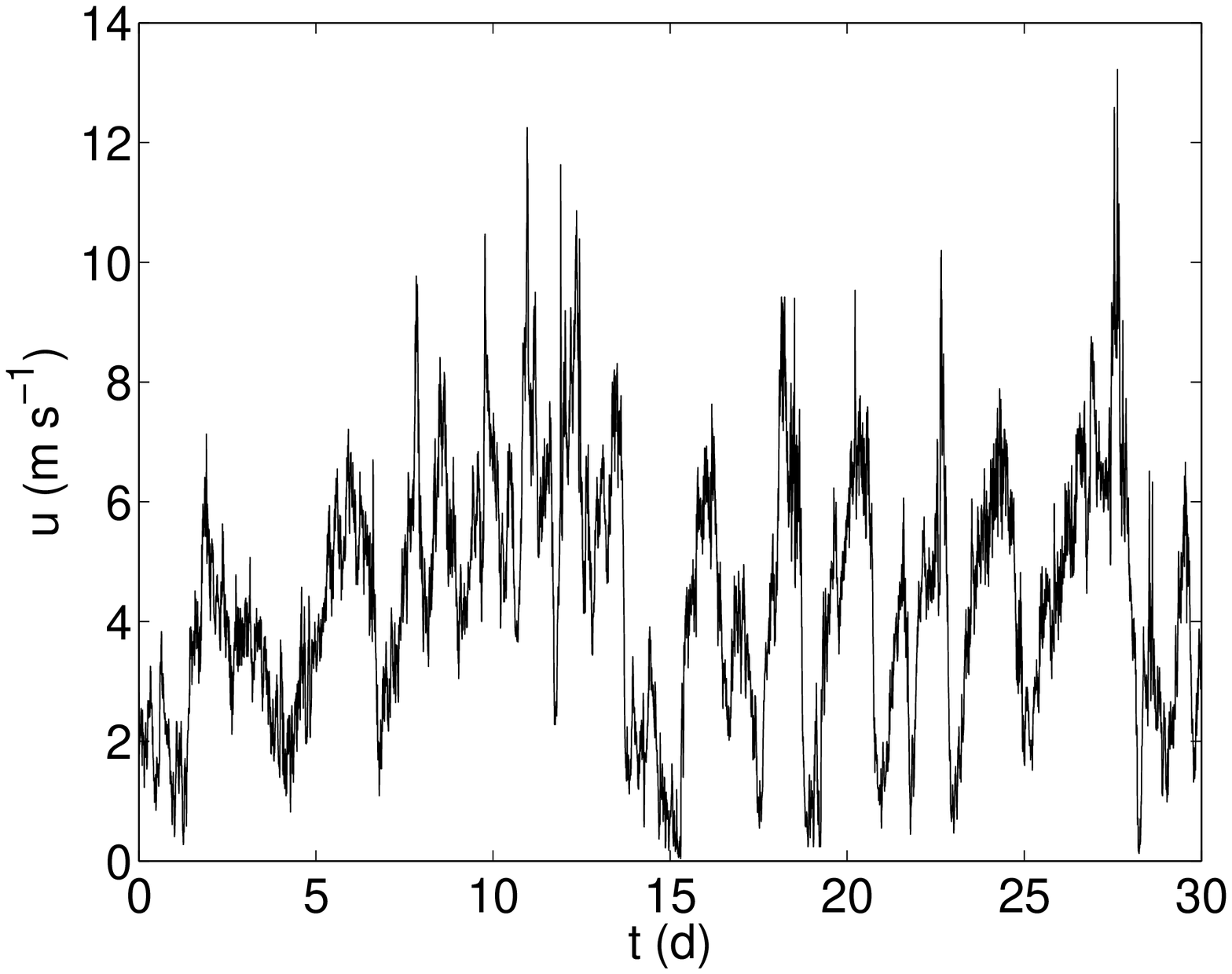}
\includegraphics[width=2.3in]{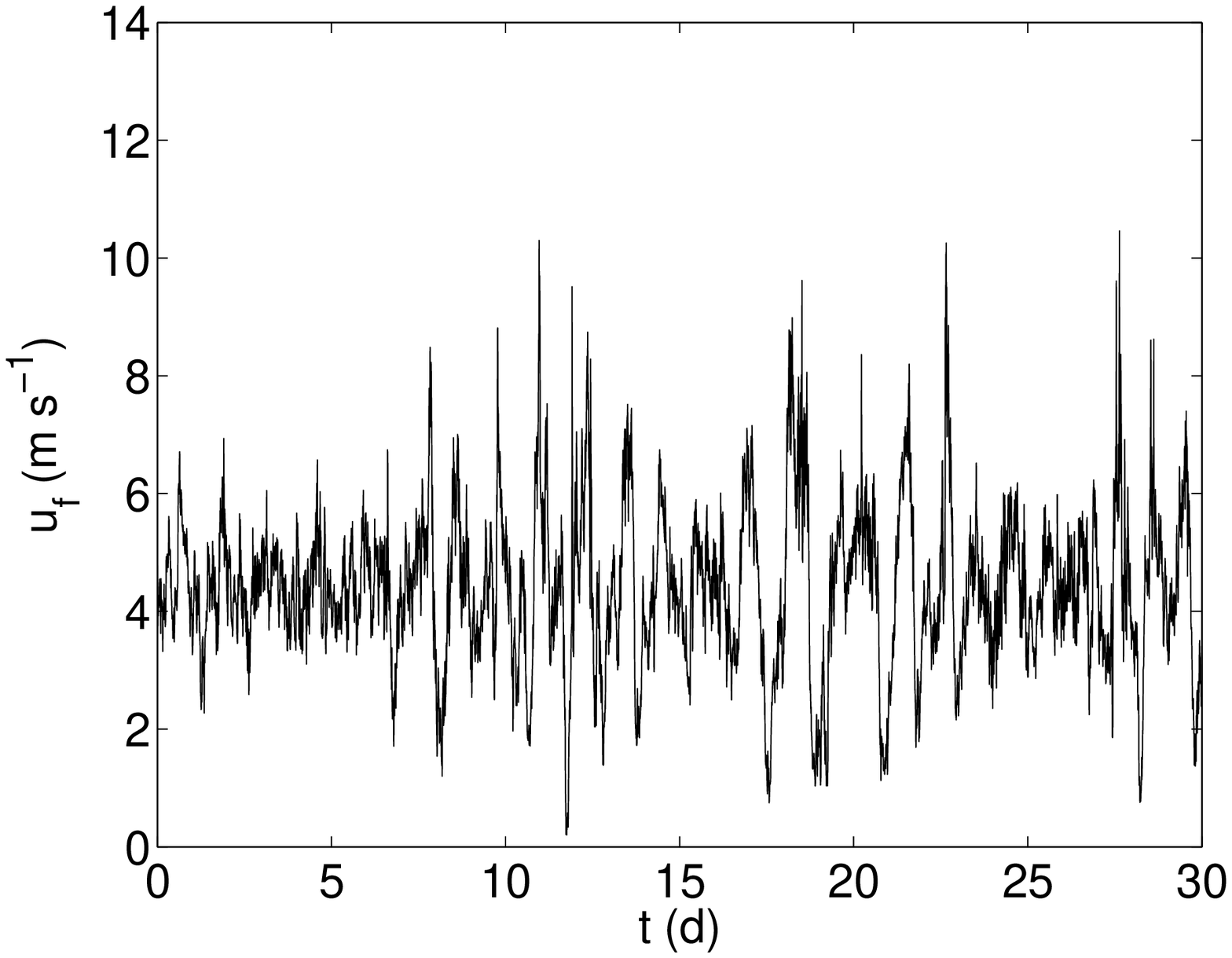}
\includegraphics[width=2.3in]{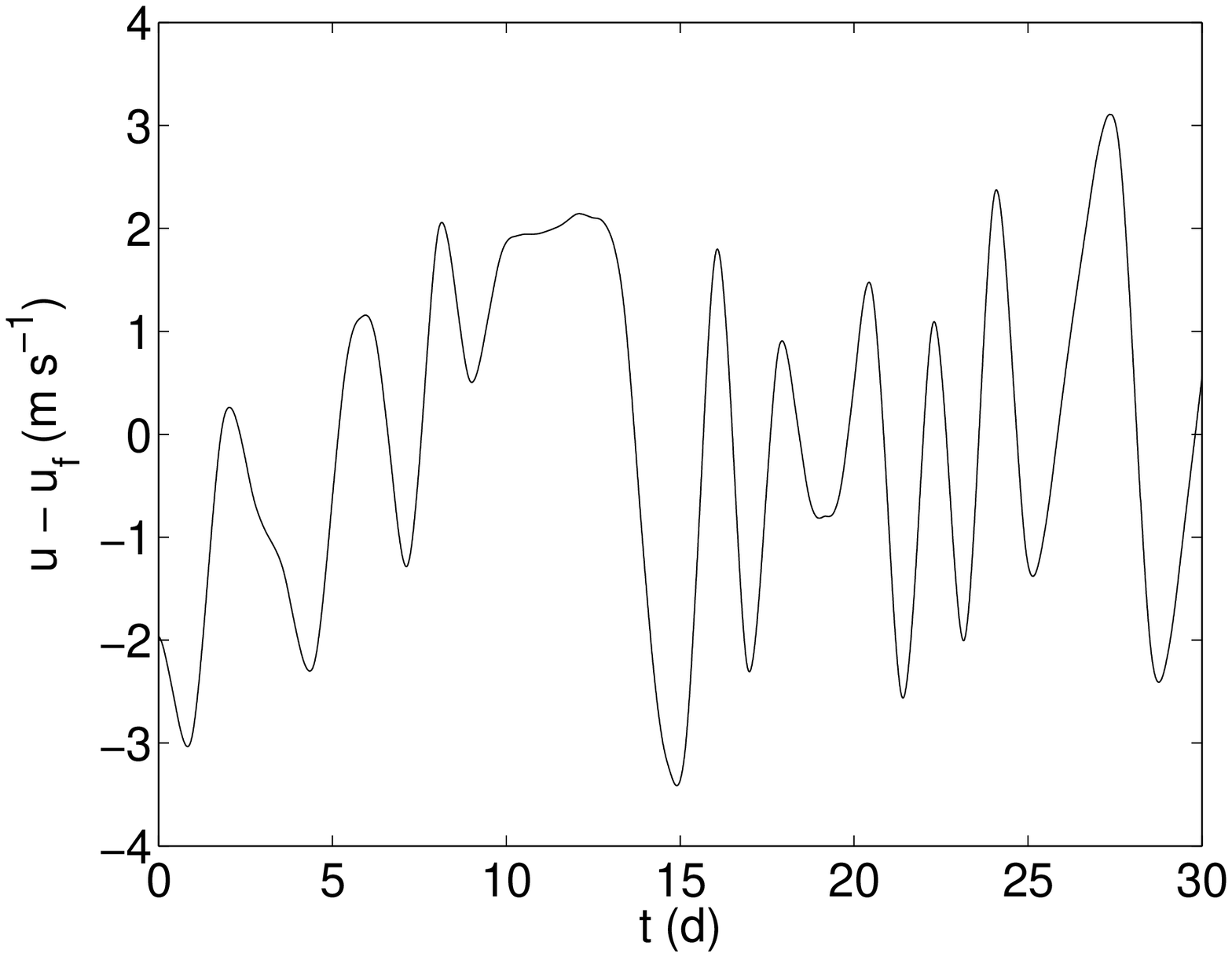}
\includegraphics[width=2.3in]{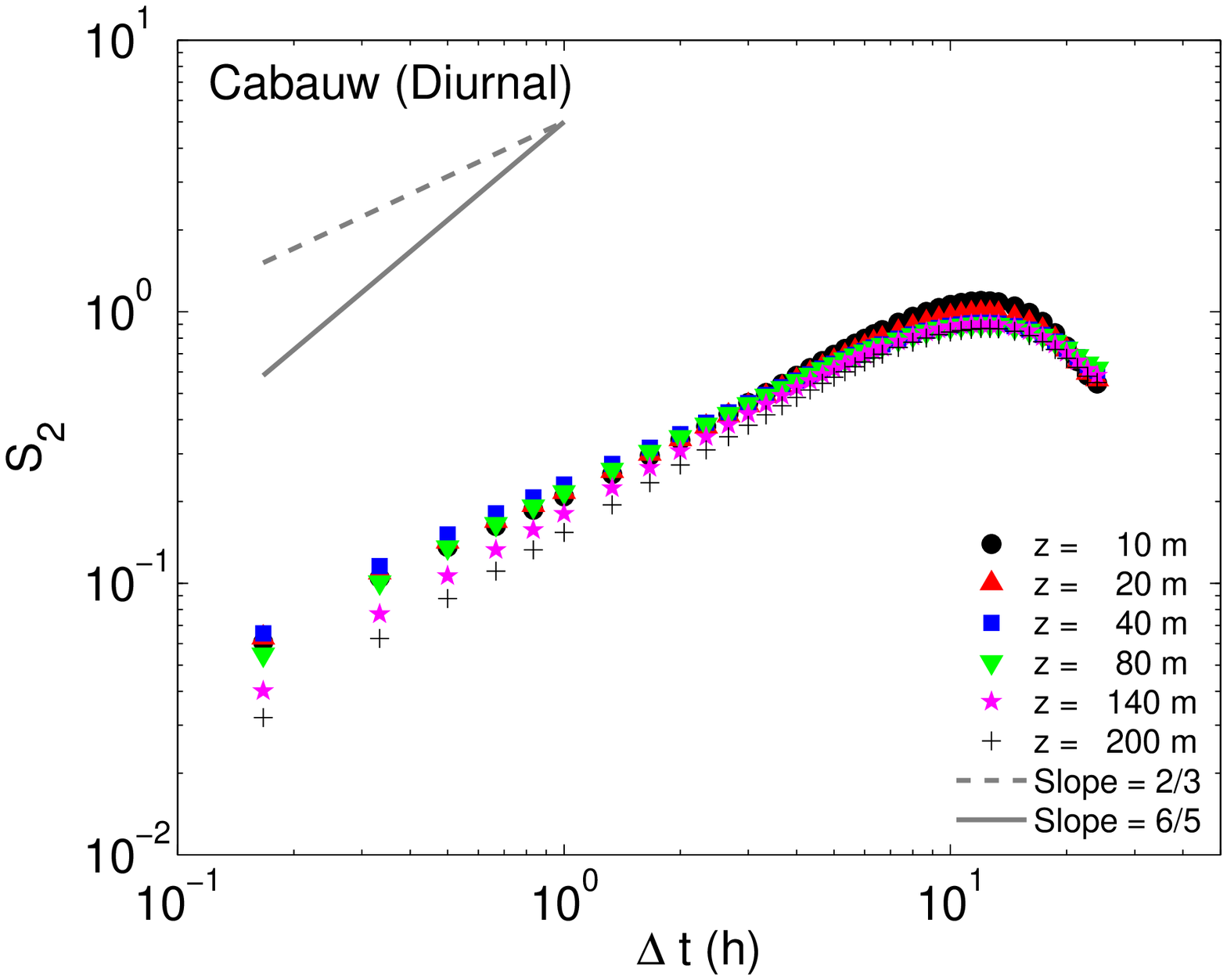}
\includegraphics[width=2.3in]{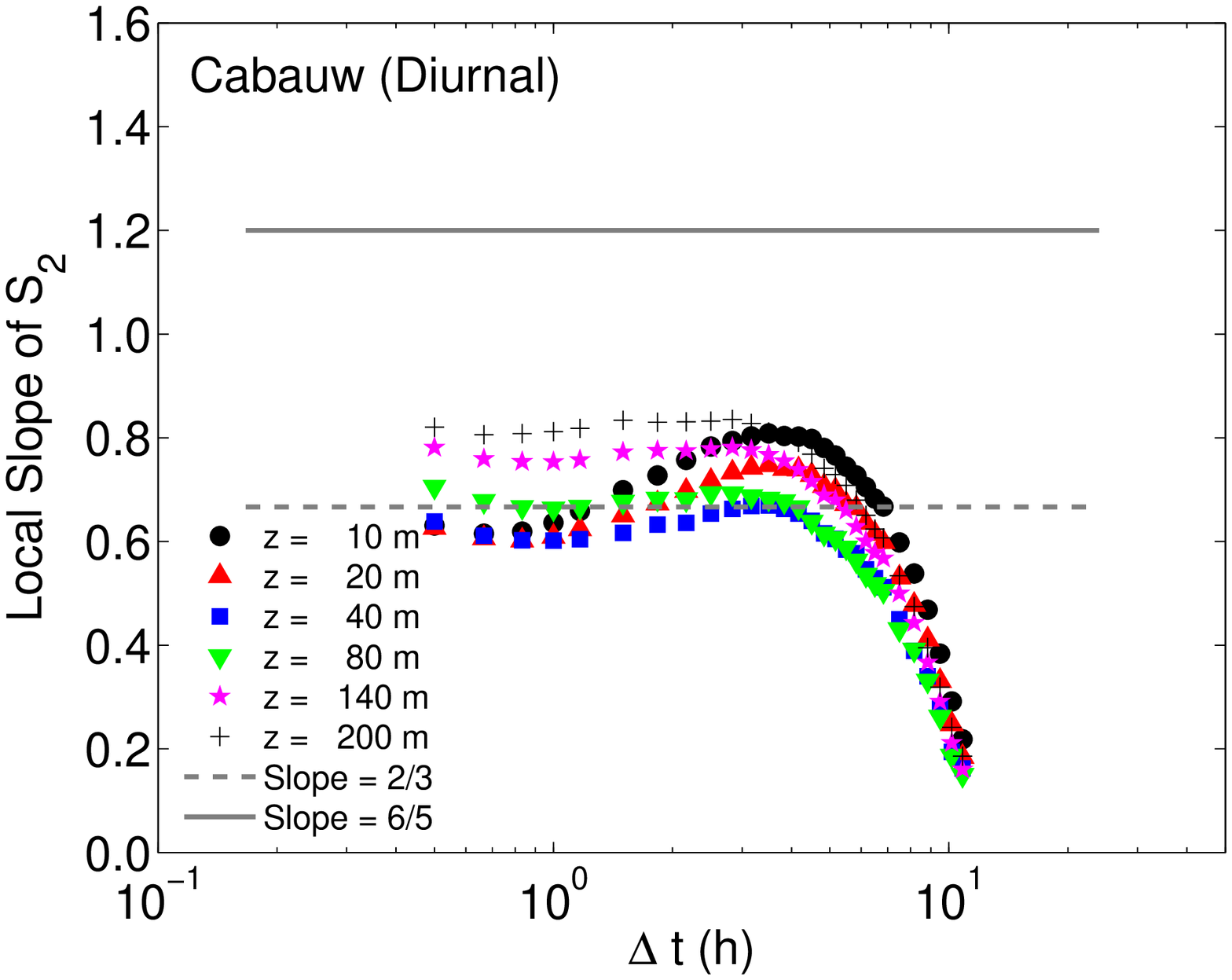}
\includegraphics[width=2.3in]{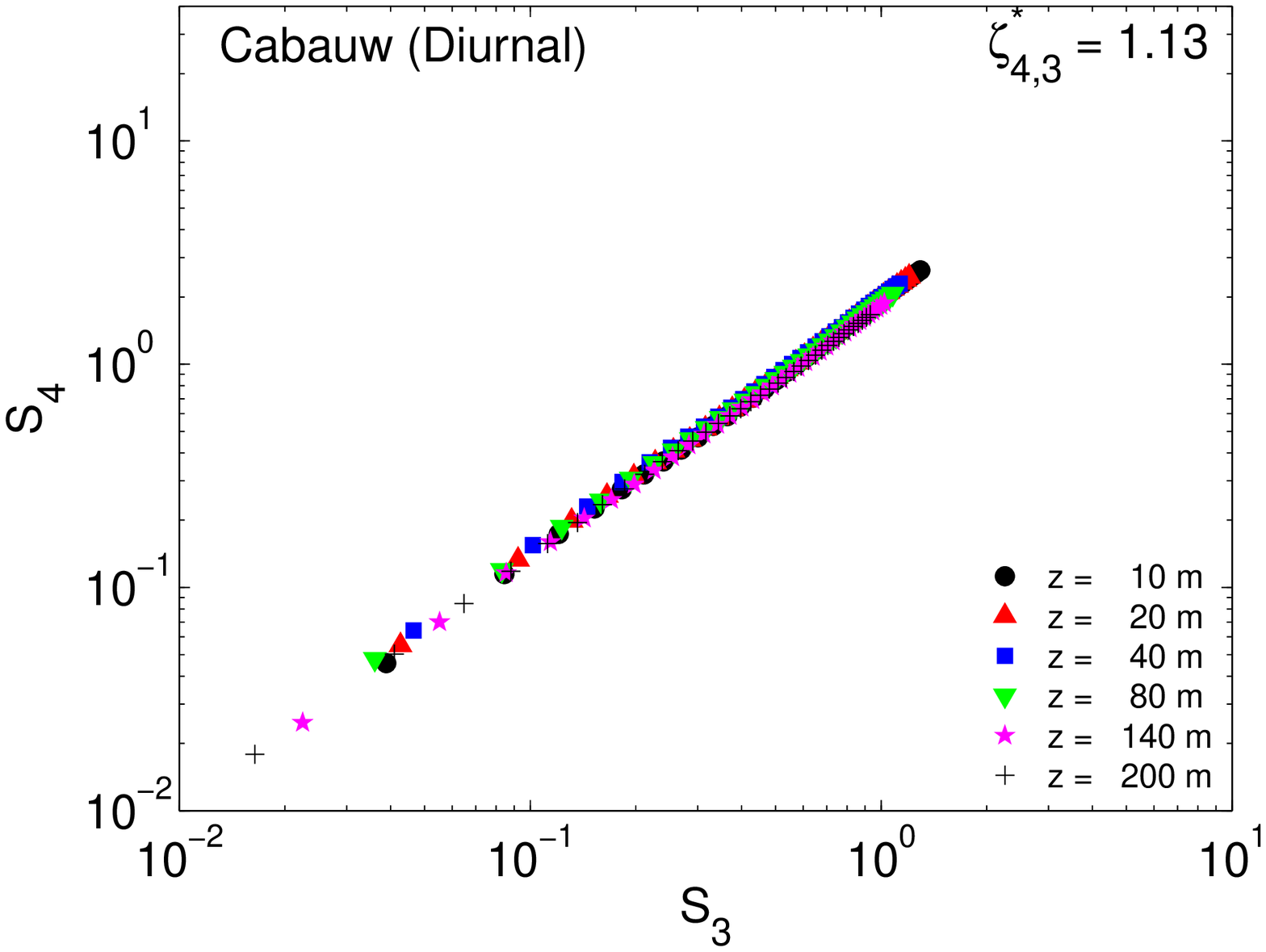}
\caption{\label{Detrend} (Color Online) Top panel plots provide an illustration of the filtering of diurnal cycles from an observed wind time-series. One month-long wind data ($u$) from the Cabauw tower is shown in the top-left panel. The top-middle panel shows the wavelet-based (high-pass) filtered series ($u_f$). The difference between the original and filtered series ($u - u_f$) is shown in the top-right panel. The plots in the bottom panel utilize 170 months of wind speed observations from the Cabauw tower. Prior to the structure function analyses, wavelet-based filtering has been employed as illustrated in the top panel. The second-order structure functions and their corresponding local slopes are shown in the bottom-left and bottom-middle panels, respectively. The dashed and solid lines represent slopes corresponding to Kolmogorov's and Bolgiano's hypotheses, respectively. The variation of the fourth-order structure function with respect to the third-order structure function is shown in bottom-right panel. The relative exponent $\zeta^{*}_{p,q}$ is reported on the top right corner of this plot.}
\end{figure*}

In this appendix, we investigate if the diurnal cycles have any impact on the scaling exponents in the mesoscale regime. For this task, we utilize discrete wavelet transform (Symmlet-8 wavelet) with a filter-scale of 21.33 h ($\approx 2^7 \times 10$ min), following Basu et al.~\cite{basu2006revisiting}. In the top-panel of Fig.~\ref{Detrend}, an illustration of the filtering approach is provided. In the bottom panel of this figure, various scaling statistics based on the filtered wind speed data from the Cabauw tower are shown. These plots can be compared against their unfiltered counter-parts reported in Fig.~\ref{SF2-Cabauw} and Fig.~\ref{ESS}. Clearly, the diurnal cycles have insignificant impact on the reported scaling exponents for $\Delta t \le 6$ h.

\section{Estimation of Stratification in the Surface Layers over the WTM Region}
\label{stratification}
\begin{figure}[htpb]
\includegraphics[width=3in]{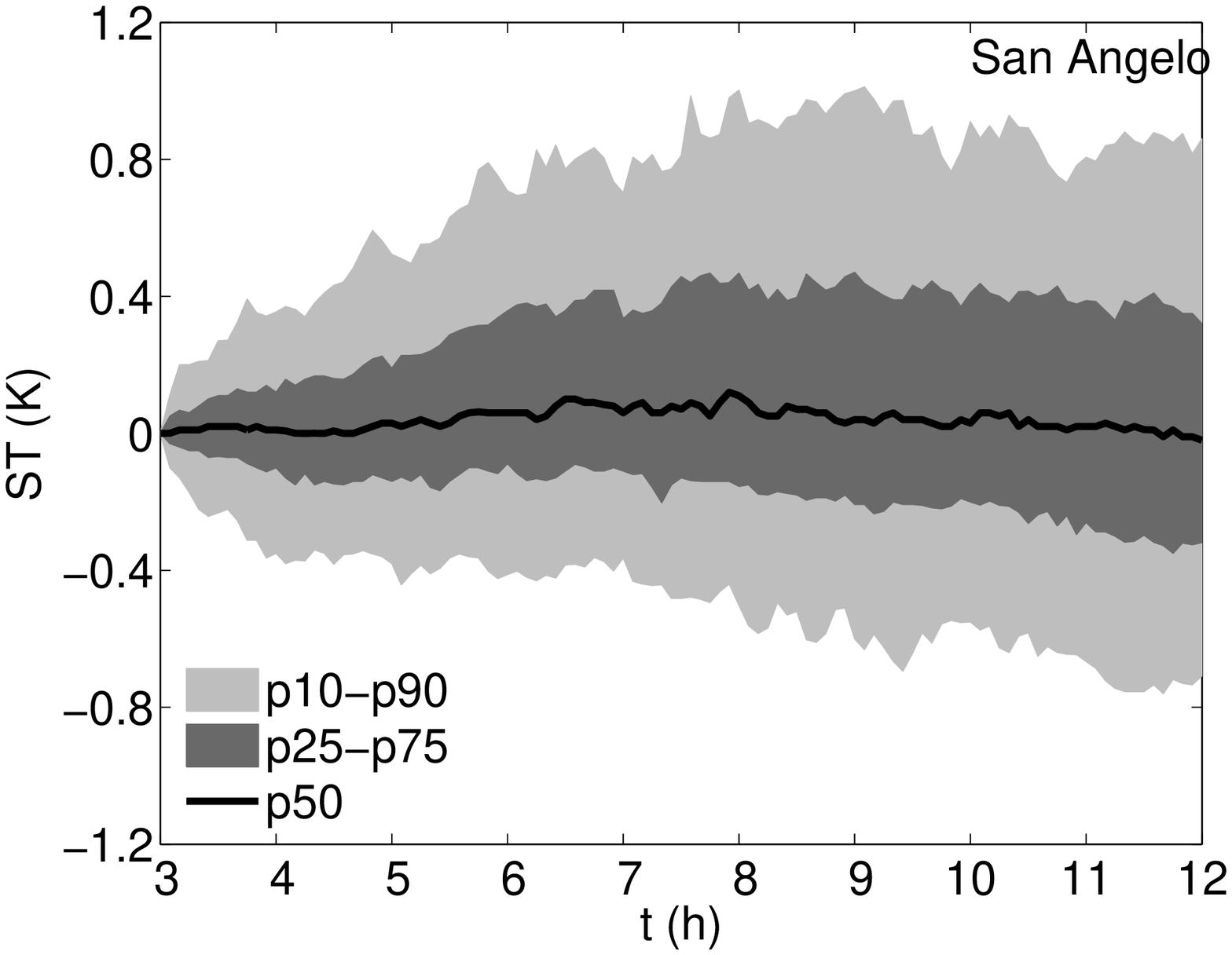}
\includegraphics[width=3in]{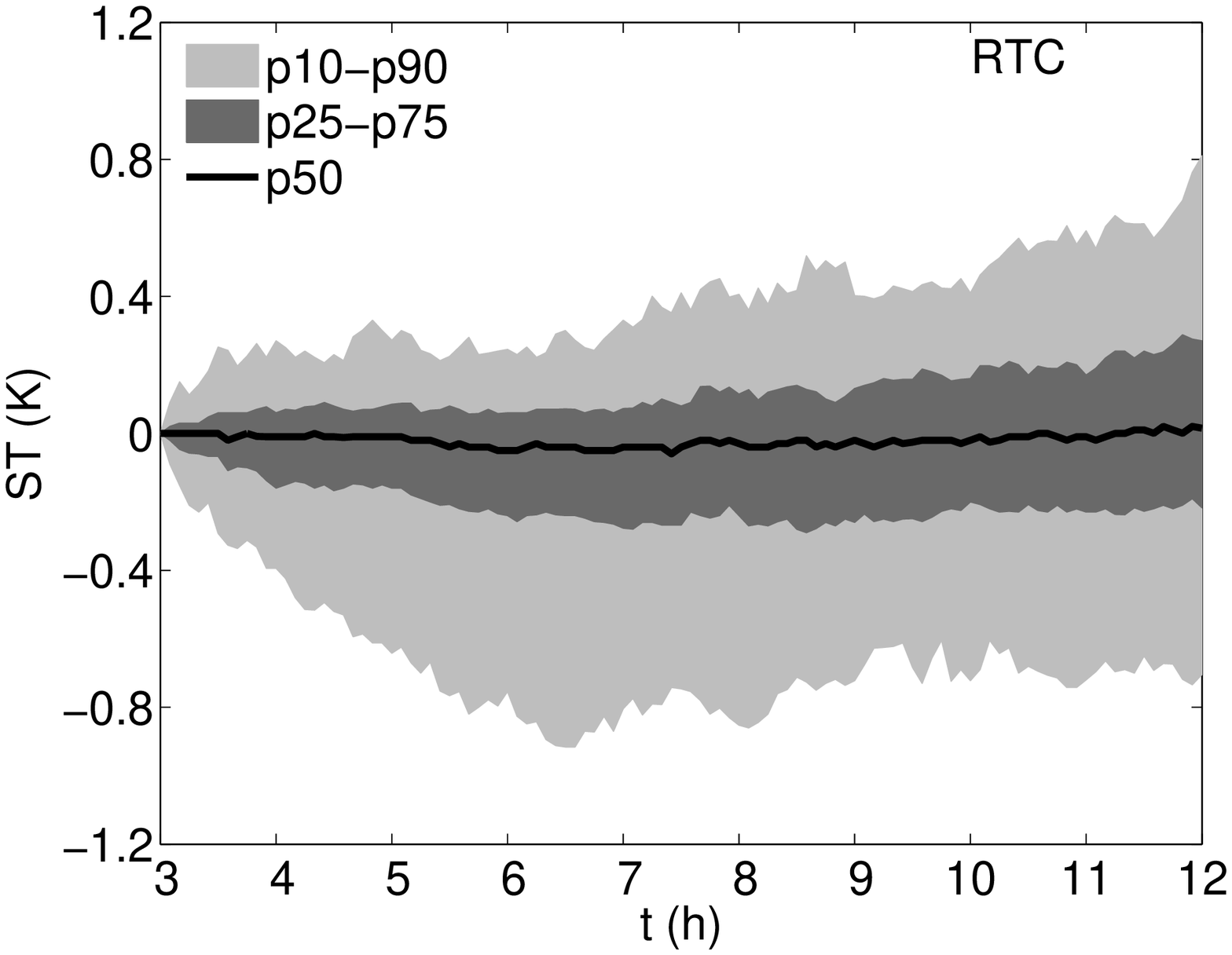}
\caption{\label{dCR} Time-series plots of stratification ($ST$) during nighttime periods. Eight months of observational data from the San Angelo (top panel) and RTC (bottom panel) mesonets are utilized. Time is in UTC. The solid black
lines, dark grey shaded areas, and the light grey areas correspond to the medians, 25\textsuperscript{th}--75\textsuperscript{th} percentile ranges, and 10\textsuperscript{th}--90\textsuperscript{th} percentile ranges, respectively.}
\end{figure}

Two 10 m tall mesonet stations are located in close proximity of the sodars at San Angelo and RTC. Air temperature (5 min average) at heights 2 m and 9 m (AGL) are available from both these stations. We first calculate the temporal evolution of nocturnal cooling as follows: 
\begin{equation}
\Delta\Theta_z(t) = \Theta_z(t) - \Theta_z(t = t_0)
\end{equation} 
where $\Theta_z$ is the mean potential temperature at height $z$. The initial time ($t_0$) is taken as 9 pm CST (i.e., 3 am UTC). Next, we estimate stratification in the 2 m -- 9 m layer via the following relationship:  
\begin{equation}
ST(t) = \Delta\Theta_{z = 9 \mbox{ m}}(t) - \Delta\Theta_{z = 2 \mbox{ m}}(t)
\end{equation}
From overall flux-divergence consideration, more negative values of $ST(t)$ signify stronger stratification. In contrast, small positive values denote weakly unstable condition. Based on 8 months of data (i.e., 243 nocturnal $ST$ time-series), Fig.~\ref{dCR} is created. Clearly, the stratification level is much stronger at RTC than at San Angelo. In other words, Fig.~\ref{dCR} provides further evidence to our earlier conjecture that the buoyancy effects are more dominant at RTC in comparison to San Angelo. 

\begin{acknowledgments}
The authors are grateful to West Texas Mesonet for sending us the sodar datasets. SB thanks Fred Bosveld (Royal Netherlands Meteorological Institute) and BMU (Bundesministerium f\"{u}r Umwelt, Federal Ministry for the Environment, Nature Conservation and Nuclear Safety) for granting him access to the Cabauw datasets (CESAR database), and FINO 1 database, respectively. Special thanks go to Bert Holtslag (Wageningen University) for providing valuable insight about the Cabauw site and relevant literature. This research was partially supported by the U.S. Department of Energy (grant \# DE-EE0004420; subcontract from AWS Truepower) and the National Science Foundation (grant \# AGS-1122315). Any opinions, findings and conclusions or recommendations expressed in this material are those of the authors and do not necessarily reflect the views of the U.S. Department of Energy or the National Science Foundation.
\end{acknowledgments}


\begin{thebibliography}{60}%
\makeatletter
\providecommand \@ifxundefined [1]{%
 \@ifx{#1\undefined}
}%
\providecommand \@ifnum [1]{%
 \ifnum #1\expandafter \@firstoftwo
 \else \expandafter \@secondoftwo
 \fi
}%
\providecommand \@ifx [1]{%
 \ifx #1\expandafter \@firstoftwo
 \else \expandafter \@secondoftwo
 \fi
}%
\providecommand \natexlab [1]{#1}%
\providecommand \enquote  [1]{``#1''}%
\providecommand \bibnamefont  [1]{#1}%
\providecommand \bibfnamefont [1]{#1}%
\providecommand \citenamefont [1]{#1}%
\providecommand \href@noop [0]{\@secondoftwo}%
\providecommand \href [0]{\begingroup \@sanitize@url \@href}%
\providecommand \@href[1]{\@@startlink{#1}\@@href}%
\providecommand \@@href[1]{\endgroup#1\@@endlink}%
\providecommand \@sanitize@url [0]{\catcode `\\12\catcode `\$12\catcode
  `\&12\catcode `\#12\catcode `\^12\catcode `\_12\catcode `\%12\relax}%
\providecommand \@@startlink[1]{}%
\providecommand \@@endlink[0]{}%
\providecommand \url  [0]{\begingroup\@sanitize@url \@url }%
\providecommand \@url [1]{\endgroup\@href {#1}{\urlprefix }}%
\providecommand \urlprefix  [0]{URL }%
\providecommand \Eprint [0]{\href }%
\providecommand \doibase [0]{http://dx.doi.org/}%
\providecommand \selectlanguage [0]{\@gobble}%
\providecommand \bibinfo  [0]{\@secondoftwo}%
\providecommand \bibfield  [0]{\@secondoftwo}%
\providecommand \translation [1]{[#1]}%
\providecommand \BibitemOpen [0]{}%
\providecommand \bibitemStop [0]{}%
\providecommand \bibitemNoStop [0]{.\EOS\space}%
\providecommand \EOS [0]{\spacefactor3000\relax}%
\providecommand \BibitemShut  [1]{\csname bibitem#1\endcsname}%
\let\auto@bib@innerbib\@empty
\bibitem [{Note1()}]{Note1}%
  \BibitemOpen
  \bibinfo {note} {During nighttime, the depth of the boundary layer ($h$) can
  be very shallow; $h \ll 100$ m. In contrast, during the daytime (over land),
  $h$ can be on the order of 2--3 km.}\BibitemShut {Stop}%
\bibitem [{\citenamefont {Stull}(1988)}]{stul88}%
  \BibitemOpen
  \bibfield  {author} {\bibinfo {author} {\bibfnamefont {R.~B.}\ \bibnamefont
  {Stull}},\ }\href@noop {} {\emph {\bibinfo {title} {An Introduction to
  Boundary Layer Meteorology}}}\ (\bibinfo  {publisher} {Kluwer Academic
  Publishers},\ \bibinfo {address} {Dordrecht, The Netherlands},\ \bibinfo
  {year} {1988})\ p.\ \bibinfo {pages} {670 pp.}\BibitemShut {Stop}%
\bibitem [{\citenamefont {Garratt}(1992)}]{garr92}%
  \BibitemOpen
  \bibfield  {author} {\bibinfo {author} {\bibfnamefont {J.~R.}\ \bibnamefont
  {Garratt}},\ }\href@noop {} {\emph {\bibinfo {title} {The Atmospheric
  Boundary Layer}}}\ (\bibinfo  {publisher} {Cambridge University Press},\
  \bibinfo {address} {Cambridge, UK},\ \bibinfo {year} {1992})\ p.\ \bibinfo
  {pages} {316 pp.}\BibitemShut {Stop}%
\bibitem [{\citenamefont {Van~der Hoven}(1957)}]{van1957power}%
  \BibitemOpen
  \bibfield  {author} {\bibinfo {author} {\bibfnamefont {I.}~\bibnamefont
  {Van~der Hoven}},\ }\href@noop {} {\bibfield  {journal} {\bibinfo  {journal}
  {J. Meteorol.}\ }\textbf {\bibinfo {volume} {14}},\ \bibinfo {pages} {160}
  (\bibinfo {year} {1957})}\BibitemShut {NoStop}%
\bibitem [{\citenamefont {Kolesnikova}\ and\ \citenamefont
  {Monin}(1968)}]{kolesnikova1968spectra}%
  \BibitemOpen
  \bibfield  {author} {\bibinfo {author} {\bibfnamefont {V.~N.}\ \bibnamefont
  {Kolesnikova}}\ and\ \bibinfo {author} {\bibfnamefont {A.~S.}\ \bibnamefont
  {Monin}},\ }\href@noop {} {\bibfield  {journal} {\bibinfo  {journal}
  {Meteorologicheskie Issledovaniya}\ }\textbf {\bibinfo {volume} {16}},\
  \bibinfo {pages} {30} (\bibinfo {year} {1968})}\BibitemShut {NoStop}%
\bibitem [{\citenamefont {Lyons}(1975)}]{lyons1975mesoscale}%
  \BibitemOpen
  \bibfield  {author} {\bibinfo {author} {\bibfnamefont {T.~J.}\ \bibnamefont
  {Lyons}},\ }\href@noop {} {\bibfield  {journal} {\bibinfo  {journal} {Q. J.
  Roy. Meteorol. Soc.}\ }\textbf {\bibinfo {volume} {101}},\ \bibinfo {pages}
  {901} (\bibinfo {year} {1975})}\BibitemShut {NoStop}%
\bibitem [{\citenamefont {Ishida}(1990)}]{ishida1990seasonal}%
  \BibitemOpen
  \bibfield  {author} {\bibinfo {author} {\bibfnamefont {H.}~\bibnamefont
  {Ishida}},\ }\href@noop {} {\bibfield  {journal} {\bibinfo  {journal}
  {Bound.-Lay. Meteorol.}\ }\textbf {\bibinfo {volume} {52}},\ \bibinfo {pages}
  {335} (\bibinfo {year} {1990})}\BibitemShut {NoStop}%
\bibitem [{\citenamefont {Lars{\'e}n}\ \emph {et~al.}(2011)\citenamefont
  {Lars{\'e}n}, \citenamefont {Larsen},\ and\ \citenamefont
  {Badger}}]{larsen2011case}%
  \BibitemOpen
  \bibfield  {author} {\bibinfo {author} {\bibfnamefont {X.~G.}\ \bibnamefont
  {Lars{\'e}n}}, \bibinfo {author} {\bibfnamefont {S.}~\bibnamefont {Larsen}},
  \ and\ \bibinfo {author} {\bibfnamefont {M.}~\bibnamefont {Badger}},\
  }\href@noop {} {\bibfield  {journal} {\bibinfo  {journal} {Q. J. Roy.
  Meteorol. Soc.}\ }\textbf {\bibinfo {volume} {137}},\ \bibinfo {pages} {264}
  (\bibinfo {year} {2011})}\BibitemShut {NoStop}%
\bibitem [{\citenamefont {Chambers}\ and\ \citenamefont
  {Antonia}(1984)}]{chambers1984atmospheric}%
  \BibitemOpen
  \bibfield  {author} {\bibinfo {author} {\bibfnamefont {A.~J.}\ \bibnamefont
  {Chambers}}\ and\ \bibinfo {author} {\bibfnamefont {R.~A.}\ \bibnamefont
  {Antonia}},\ }\href@noop {} {\bibfield  {journal} {\bibinfo  {journal}
  {Bound.-Lay. Meteorol.}\ }\textbf {\bibinfo {volume} {28}},\ \bibinfo {pages}
  {343} (\bibinfo {year} {1984})}\BibitemShut {NoStop}%
\bibitem [{\citenamefont {Sreenivasan}\ and\ \citenamefont
  {Kailasnath}(1993)}]{sreenivasan1993update}%
  \BibitemOpen
  \bibfield  {author} {\bibinfo {author} {\bibfnamefont {K.}~\bibnamefont
  {Sreenivasan}}\ and\ \bibinfo {author} {\bibfnamefont {P.}~\bibnamefont
  {Kailasnath}},\ }\href@noop {} {\bibfield  {journal} {\bibinfo  {journal}
  {Phys. Fluids A}\ }\textbf {\bibinfo {volume} {5}},\ \bibinfo {pages} {512}
  (\bibinfo {year} {1993})}\BibitemShut {NoStop}%
\bibitem [{\citenamefont {Praskovsky}\ and\ \citenamefont
  {Oncley}(1997)}]{praskovsky1997comprehensive}%
  \BibitemOpen
  \bibfield  {author} {\bibinfo {author} {\bibfnamefont {A.}~\bibnamefont
  {Praskovsky}}\ and\ \bibinfo {author} {\bibfnamefont {S.}~\bibnamefont
  {Oncley}},\ }\href@noop {} {\bibfield  {journal} {\bibinfo  {journal} {Fluid
  Dyn. Res.}\ }\textbf {\bibinfo {volume} {21}},\ \bibinfo {pages} {331}
  (\bibinfo {year} {1997})}\BibitemShut {NoStop}%
\bibitem [{\citenamefont {Kurien}\ \emph {et~al.}(2000)\citenamefont {Kurien},
  \citenamefont {L'vov}, \citenamefont {Procaccia},\ and\ \citenamefont
  {Sreenivasan}}]{kurien2000scaling}%
  \BibitemOpen
  \bibfield  {author} {\bibinfo {author} {\bibfnamefont {S.}~\bibnamefont
  {Kurien}}, \bibinfo {author} {\bibfnamefont {V.~S.}\ \bibnamefont {L'vov}},
  \bibinfo {author} {\bibfnamefont {I.}~\bibnamefont {Procaccia}}, \ and\
  \bibinfo {author} {\bibfnamefont {K.}~\bibnamefont {Sreenivasan}},\
  }\href@noop {} {\bibfield  {journal} {\bibinfo  {journal} {Phys. Rev. E}\
  }\textbf {\bibinfo {volume} {61}},\ \bibinfo {pages} {407} (\bibinfo {year}
  {2000})}\BibitemShut {NoStop}%
\bibitem [{\citenamefont {Basu}\ \emph {et~al.}(2007)\citenamefont {Basu},
  \citenamefont {Foufoula-Georgiou}, \citenamefont {Lashermes},\ and\
  \citenamefont {Arn{\'e}odo}}]{basu2007estimating}%
  \BibitemOpen
  \bibfield  {author} {\bibinfo {author} {\bibfnamefont {S.}~\bibnamefont
  {Basu}}, \bibinfo {author} {\bibfnamefont {E.}~\bibnamefont
  {Foufoula-Georgiou}}, \bibinfo {author} {\bibfnamefont {B.}~\bibnamefont
  {Lashermes}}, \ and\ \bibinfo {author} {\bibfnamefont {A.}~\bibnamefont
  {Arn{\'e}odo}},\ }\href@noop {} {\bibfield  {journal} {\bibinfo  {journal}
  {Phys. Fluids}\ }\textbf {\bibinfo {volume} {19}},\ \bibinfo {pages} {115102}
  (\bibinfo {year} {2007})}\BibitemShut {NoStop}%
\bibitem [{\citenamefont {Morales}\ \emph {et~al.}(2012)\citenamefont
  {Morales}, \citenamefont {W{\"a}chter},\ and\ \citenamefont
  {Peinke}}]{morales2012characterization}%
  \BibitemOpen
  \bibfield  {author} {\bibinfo {author} {\bibfnamefont {A.}~\bibnamefont
  {Morales}}, \bibinfo {author} {\bibfnamefont {M.}~\bibnamefont
  {W{\"a}chter}}, \ and\ \bibinfo {author} {\bibfnamefont {J.}~\bibnamefont
  {Peinke}},\ }\href@noop {} {\bibfield  {journal} {\bibinfo  {journal} {Wind
  Energy}\ }\textbf {\bibinfo {volume} {15}},\ \bibinfo {pages} {391} (\bibinfo
  {year} {2012})}\BibitemShut {NoStop}%
\bibitem [{Note2()}]{Note2}%
  \BibitemOpen
  \bibinfo {note} {Quite a few present-day authors inappropriately use the
  terms anomalous scaling and multifractality interchangeably. For specific
  cases, the multifractal formalism is a plausible way of explaining anomalous
  scaling behavior. However, it is not always applicable.}\BibitemShut {Stop}%
\bibitem [{\citenamefont {Lauren}\ \emph {et~al.}(1999)\citenamefont {Lauren},
  \citenamefont {Menabde}, \citenamefont {Seed},\ and\ \citenamefont
  {Austin}}]{lauren1999characterisation}%
  \BibitemOpen
  \bibfield  {author} {\bibinfo {author} {\bibfnamefont {M.~K.}\ \bibnamefont
  {Lauren}}, \bibinfo {author} {\bibfnamefont {M.}~\bibnamefont {Menabde}},
  \bibinfo {author} {\bibfnamefont {A.~W.}\ \bibnamefont {Seed}}, \ and\
  \bibinfo {author} {\bibfnamefont {G.~L.}\ \bibnamefont {Austin}},\
  }\href@noop {} {\bibfield  {journal} {\bibinfo  {journal} {Bound.-Lay.
  Meteorol.}\ }\textbf {\bibinfo {volume} {90}},\ \bibinfo {pages} {21}
  (\bibinfo {year} {1999})}\BibitemShut {NoStop}%
\bibitem [{\citenamefont {Lauren}\ \emph {et~al.}(2001)\citenamefont {Lauren},
  \citenamefont {Menabde},\ and\ \citenamefont {Austin}}]{lauren2001analysis}%
  \BibitemOpen
  \bibfield  {author} {\bibinfo {author} {\bibfnamefont {M.~K.}\ \bibnamefont
  {Lauren}}, \bibinfo {author} {\bibfnamefont {M.}~\bibnamefont {Menabde}}, \
  and\ \bibinfo {author} {\bibfnamefont {G.~L.}\ \bibnamefont {Austin}},\
  }\href@noop {} {\bibfield  {journal} {\bibinfo  {journal} {Bound.-Lay.
  Meteorol.}\ }\textbf {\bibinfo {volume} {100}},\ \bibinfo {pages} {263}
  (\bibinfo {year} {2001})}\BibitemShut {NoStop}%
\bibitem [{\citenamefont {Govindan}\ and\ \citenamefont
  {Kantz}(2004)}]{govindan2004long}%
  \BibitemOpen
  \bibfield  {author} {\bibinfo {author} {\bibfnamefont {R.~B.}\ \bibnamefont
  {Govindan}}\ and\ \bibinfo {author} {\bibfnamefont {H.}~\bibnamefont
  {Kantz}},\ }\href@noop {} {\bibfield  {journal} {\bibinfo  {journal}
  {Europhys. Lett.}\ }\textbf {\bibinfo {volume} {68}},\ \bibinfo {pages} {184}
  (\bibinfo {year} {2004})}\BibitemShut {NoStop}%
\bibitem [{\citenamefont {Kavasseri}\ and\ \citenamefont
  {Nagarajan}(2005)}]{kavasseri2005multifractal}%
  \BibitemOpen
  \bibfield  {author} {\bibinfo {author} {\bibfnamefont {R.~G.}\ \bibnamefont
  {Kavasseri}}\ and\ \bibinfo {author} {\bibfnamefont {R.}~\bibnamefont
  {Nagarajan}},\ }\href@noop {} {\bibfield  {journal} {\bibinfo  {journal}
  {Chaos, Solitons \& Fractals}\ }\textbf {\bibinfo {volume} {24}},\ \bibinfo
  {pages} {165} (\bibinfo {year} {2005})}\BibitemShut {NoStop}%
\bibitem [{\citenamefont {Ko{\c{c}}ak}(2009)}]{koccak2009examination}%
  \BibitemOpen
  \bibfield  {author} {\bibinfo {author} {\bibfnamefont {K.}~\bibnamefont
  {Ko{\c{c}}ak}},\ }\href@noop {} {\bibfield  {journal} {\bibinfo  {journal}
  {Energy}\ }\textbf {\bibinfo {volume} {34}},\ \bibinfo {pages} {1980}
  (\bibinfo {year} {2009})}\BibitemShut {NoStop}%
\bibitem [{\citenamefont {Muzy}\ \emph {et~al.}(2010)\citenamefont {Muzy},
  \citenamefont {Ba{\"\i}le},\ and\ \citenamefont
  {Poggi}}]{muzy2010intermittency}%
  \BibitemOpen
  \bibfield  {author} {\bibinfo {author} {\bibfnamefont {J.-F.}\ \bibnamefont
  {Muzy}}, \bibinfo {author} {\bibfnamefont {R.}~\bibnamefont {Ba{\"\i}le}}, \
  and\ \bibinfo {author} {\bibfnamefont {P.}~\bibnamefont {Poggi}},\
  }\href@noop {} {\bibfield  {journal} {\bibinfo  {journal} {Phys. Rev. E}\
  }\textbf {\bibinfo {volume} {81}},\ \bibinfo {pages} {056308} (\bibinfo
  {year} {2010})}\BibitemShut {NoStop}%
\bibitem [{\citenamefont {Ba{\"\i}le}\ and\ \citenamefont
  {Muzy}(2010)}]{baile2010spatial}%
  \BibitemOpen
  \bibfield  {author} {\bibinfo {author} {\bibfnamefont {R.}~\bibnamefont
  {Ba{\"\i}le}}\ and\ \bibinfo {author} {\bibfnamefont {J.-F.}\ \bibnamefont
  {Muzy}},\ }\href@noop {} {\bibfield  {journal} {\bibinfo  {journal} {Phys.
  Rev. Lett.}\ }\textbf {\bibinfo {volume} {105}},\ \bibinfo {pages} {254501}
  (\bibinfo {year} {2010})}\BibitemShut {NoStop}%
\bibitem [{\citenamefont {Telesca}\ and\ \citenamefont
  {Lovallo}(2011)}]{telesca2011analysis}%
  \BibitemOpen
  \bibfield  {author} {\bibinfo {author} {\bibfnamefont {L.}~\bibnamefont
  {Telesca}}\ and\ \bibinfo {author} {\bibfnamefont {M.}~\bibnamefont
  {Lovallo}},\ }\href@noop {} {\bibfield  {journal} {\bibinfo  {journal} {J.
  Stat. Mech.: Theory and Experiment}\ }\textbf {\bibinfo {volume} {2011}},\
  \bibinfo {pages} {P07001} (\bibinfo {year} {2011})}\BibitemShut {NoStop}%
\bibitem [{\citenamefont {Liu}\ and\ \citenamefont
  {Hu}(2013)}]{liu2013cascade}%
  \BibitemOpen
  \bibfield  {author} {\bibinfo {author} {\bibfnamefont {L.}~\bibnamefont
  {Liu}}\ and\ \bibinfo {author} {\bibfnamefont {F.}~\bibnamefont {Hu}},\
  }\href@noop {} {\bibfield  {journal} {\bibinfo  {journal} {Physica A}\
  }\textbf {\bibinfo {volume} {392}},\ \bibinfo {pages} {5808} (\bibinfo {year}
  {2013})}\BibitemShut {NoStop}%
\bibitem [{\citenamefont {Neumann}\ \emph {et~al.}(2003)\citenamefont
  {Neumann}, \citenamefont {Nolopp}, \citenamefont {Strack}, \citenamefont
  {Mellinghoff}, \citenamefont {S{\"o}ker}, \citenamefont {Mittelstaedt},
  \citenamefont {Gerasch},\ and\ \citenamefont
  {Fischer}}]{neumann2003erection}%
  \BibitemOpen
  \bibfield  {author} {\bibinfo {author} {\bibfnamefont {T.}~\bibnamefont
  {Neumann}}, \bibinfo {author} {\bibfnamefont {K.}~\bibnamefont {Nolopp}},
  \bibinfo {author} {\bibfnamefont {M.}~\bibnamefont {Strack}}, \bibinfo
  {author} {\bibfnamefont {H.}~\bibnamefont {Mellinghoff}}, \bibinfo {author}
  {\bibfnamefont {H.}~\bibnamefont {S{\"o}ker}}, \bibinfo {author}
  {\bibfnamefont {E.}~\bibnamefont {Mittelstaedt}}, \bibinfo {author}
  {\bibfnamefont {W.}~\bibnamefont {Gerasch}}, \ and\ \bibinfo {author}
  {\bibfnamefont {G.}~\bibnamefont {Fischer}},\ }\href@noop {} {\bibfield
  {journal} {\bibinfo  {journal} {DEWI Mag}\ }\textbf {\bibinfo {volume}
  {23}},\ \bibinfo {pages} {32} (\bibinfo {year} {2003})}\BibitemShut {NoStop}%
\bibitem [{\citenamefont {T{\"u}rk}\ \emph {et~al.}(2008)\citenamefont
  {T{\"u}rk}, \citenamefont {Grigutsch},\ and\ \citenamefont
  {Emeis}}]{turk2008wind}%
  \BibitemOpen
  \bibfield  {author} {\bibinfo {author} {\bibfnamefont {M.}~\bibnamefont
  {T{\"u}rk}}, \bibinfo {author} {\bibfnamefont {K.}~\bibnamefont {Grigutsch}},
  \ and\ \bibinfo {author} {\bibfnamefont {S.}~\bibnamefont {Emeis}},\
  }\href@noop {} {\bibfield  {journal} {\bibinfo  {journal} {DEWI Mag}\
  }\textbf {\bibinfo {volume} {33}},\ \bibinfo {pages} {12} (\bibinfo {year}
  {2008})}\BibitemShut {NoStop}%
\bibitem [{\citenamefont {Ernst}\ and\ \citenamefont
  {Seume}(2012)}]{ernst2012investigation}%
  \BibitemOpen
  \bibfield  {author} {\bibinfo {author} {\bibfnamefont {B.}~\bibnamefont
  {Ernst}}\ and\ \bibinfo {author} {\bibfnamefont {J.~R.}\ \bibnamefont
  {Seume}},\ }\href@noop {} {\bibfield  {journal} {\bibinfo  {journal}
  {Energies}\ }\textbf {\bibinfo {volume} {5}},\ \bibinfo {pages} {3835}
  (\bibinfo {year} {2012})}\BibitemShut {NoStop}%
\bibitem [{\citenamefont {Nieuwstadt}(1978)}]{nieuwstadt1978computation}%
  \BibitemOpen
  \bibfield  {author} {\bibinfo {author} {\bibfnamefont {F.~T.~M.}\
  \bibnamefont {Nieuwstadt}},\ }\href@noop {} {\bibfield  {journal} {\bibinfo
  {journal} {Bound.-Lay. Meteorol.}\ }\textbf {\bibinfo {volume} {14}},\
  \bibinfo {pages} {235} (\bibinfo {year} {1978})}\BibitemShut {NoStop}%
\bibitem [{\citenamefont {Nieuwstadt}(1984)}]{nieuwstadt1984turbulent}%
  \BibitemOpen
  \bibfield  {author} {\bibinfo {author} {\bibfnamefont {F.~T.~M.}\
  \bibnamefont {Nieuwstadt}},\ }\href@noop {} {\bibfield  {journal} {\bibinfo
  {journal} {J. Atmos. Sci.}\ }\textbf {\bibinfo {volume} {41}},\ \bibinfo
  {pages} {2202} (\bibinfo {year} {1984})}\BibitemShut {NoStop}%
\bibitem [{\citenamefont {Beljaars}\ and\ \citenamefont
  {Holtslag}(1991)}]{beljaars1991flux}%
  \BibitemOpen
  \bibfield  {author} {\bibinfo {author} {\bibfnamefont {A.~C.~M.}\
  \bibnamefont {Beljaars}}\ and\ \bibinfo {author} {\bibfnamefont {A.~A.~M.}\
  \bibnamefont {Holtslag}},\ }\href@noop {} {\bibfield  {journal} {\bibinfo
  {journal} {J. Appl. Meteorol.}\ }\textbf {\bibinfo {volume} {30}},\ \bibinfo
  {pages} {327} (\bibinfo {year} {1991})}\BibitemShut {NoStop}%
\bibitem [{\citenamefont {Verkaik}\ and\ \citenamefont
  {Holtslag}(2007)}]{verkaik2007wind}%
  \BibitemOpen
  \bibfield  {author} {\bibinfo {author} {\bibfnamefont {J.~W.}\ \bibnamefont
  {Verkaik}}\ and\ \bibinfo {author} {\bibfnamefont {A.~A.~M.}\ \bibnamefont
  {Holtslag}},\ }\href@noop {} {\bibfield  {journal} {\bibinfo  {journal}
  {Bound.-Lay. Meteorol.}\ }\textbf {\bibinfo {volume} {122}},\ \bibinfo
  {pages} {701} (\bibinfo {year} {2007})}\BibitemShut {NoStop}%
\bibitem [{\citenamefont {Bradley}(2008)}]{bradley2008atmospheric}%
  \BibitemOpen
  \bibfield  {author} {\bibinfo {author} {\bibfnamefont {S.}~\bibnamefont
  {Bradley}},\ }\href@noop {} {\emph {\bibinfo {title} {Atmospheric acoustic
  remote sensing}}}\ (\bibinfo  {publisher} {CRC Press},\ \bibinfo {year}
  {2008})\ p.\ \bibinfo {pages} {271 pp.}\BibitemShut {Stop}%
\bibitem [{\citenamefont {Frisch}(1995)}]{frisch1995turbulence}%
  \BibitemOpen
  \bibfield  {author} {\bibinfo {author} {\bibfnamefont {U.}~\bibnamefont
  {Frisch}},\ }\href@noop {} {\emph {\bibinfo {title} {Turbulence}}}\ (\bibinfo
   {publisher} {Cambridge University Press},\ \bibinfo {year} {1995})\ p.\
  \bibinfo {pages} {296 pp.}\BibitemShut {Stop}%
\bibitem [{\citenamefont {Bohr}\ \emph {et~al.}(1998)\citenamefont {Bohr},
  \citenamefont {Jensen}, \citenamefont {Paladin},\ and\ \citenamefont
  {Vulpiani}}]{bohr1998dynamical}%
  \BibitemOpen
  \bibfield  {author} {\bibinfo {author} {\bibfnamefont {T.}~\bibnamefont
  {Bohr}}, \bibinfo {author} {\bibfnamefont {M.~H.}\ \bibnamefont {Jensen}},
  \bibinfo {author} {\bibfnamefont {G.}~\bibnamefont {Paladin}}, \ and\
  \bibinfo {author} {\bibfnamefont {A.}~\bibnamefont {Vulpiani}},\ }\href@noop
  {} {\emph {\bibinfo {title} {Dynamical Systems Approach to Turbulence}}}\
  (\bibinfo  {publisher} {Cambridge University Press},\ \bibinfo {year}
  {1998})\ p.\ \bibinfo {pages} {350 pp.}\BibitemShut {Stop}%
\bibitem [{\citenamefont {Kolmogorov}(1941)}]{kolmogorov1941local}%
  \BibitemOpen
  \bibfield  {author} {\bibinfo {author} {\bibfnamefont {A.~N.}\ \bibnamefont
  {Kolmogorov}},\ }\bibfield  {booktitle} {\emph {\bibinfo {booktitle} {Dokl.
  Akad. Nauk SSSR}},\ }\href@noop {} {\ \textbf {\bibinfo {volume} {30}},\
  \bibinfo {pages} {299} (\bibinfo {year} {1941})}\BibitemShut {NoStop}%
\bibitem [{\citenamefont {Bolgiano}(1959)}]{bolgiano1959turbulent}%
  \BibitemOpen
  \bibfield  {author} {\bibinfo {author} {\bibfnamefont {R.}~\bibnamefont
  {Bolgiano}},\ }\href@noop {} {\bibfield  {journal} {\bibinfo  {journal} {J.
  Geophys. Res.}\ }\textbf {\bibinfo {volume} {64}},\ \bibinfo {pages} {2226}
  (\bibinfo {year} {1959})}\BibitemShut {NoStop}%
\bibitem [{\citenamefont {Bolgiano}(1962)}]{bolgiano1962structure}%
  \BibitemOpen
  \bibfield  {author} {\bibinfo {author} {\bibfnamefont {R.}~\bibnamefont
  {Bolgiano}},\ }\href@noop {} {\bibfield  {journal} {\bibinfo  {journal} {J.
  Geophys. Res.}\ }\textbf {\bibinfo {volume} {67}},\ \bibinfo {pages} {3015}
  (\bibinfo {year} {1962})}\BibitemShut {NoStop}%
\bibitem [{\citenamefont {Monin}\ and\ \citenamefont
  {Yaglom}(1975)}]{monin1975}%
  \BibitemOpen
  \bibfield  {author} {\bibinfo {author} {\bibfnamefont {A.~S.}\ \bibnamefont
  {Monin}}\ and\ \bibinfo {author} {\bibfnamefont {A.~M.}\ \bibnamefont
  {Yaglom}},\ }\href@noop {} {\emph {\bibinfo {title} {Statistical Fluid
  Mechanics: Mechanics of Turbulence}}},\ Vol.~\bibinfo {volume} {2}\ (\bibinfo
   {publisher} {The MIT Press},\ \bibinfo {year} {1975})\ p.\ \bibinfo {pages}
  {874 pp.}\BibitemShut {Stop}%
\bibitem [{\citenamefont {Benzi}\ \emph {et~al.}(1994)\citenamefont {Benzi},
  \citenamefont {Tripiccione}, \citenamefont {Massaioli}, \citenamefont
  {Succi},\ and\ \citenamefont {Ciliberto}}]{benzi1994scaling}%
  \BibitemOpen
  \bibfield  {author} {\bibinfo {author} {\bibfnamefont {R.}~\bibnamefont
  {Benzi}}, \bibinfo {author} {\bibfnamefont {R.}~\bibnamefont {Tripiccione}},
  \bibinfo {author} {\bibfnamefont {F.}~\bibnamefont {Massaioli}}, \bibinfo
  {author} {\bibfnamefont {S.}~\bibnamefont {Succi}}, \ and\ \bibinfo {author}
  {\bibfnamefont {S.}~\bibnamefont {Ciliberto}},\ }\href@noop {} {\bibfield
  {journal} {\bibinfo  {journal} {Europhys.\ Lett.}\ }\textbf {\bibinfo
  {volume} {25}},\ \bibinfo {pages} {341} (\bibinfo {year} {1994})}\BibitemShut
  {NoStop}%
\bibitem [{\citenamefont {Niemela}\ \emph {et~al.}(2000)\citenamefont
  {Niemela}, \citenamefont {Skrbek}, \citenamefont {Sreenivasan},\ and\
  \citenamefont {Donnelly}}]{niemela2000turbulent}%
  \BibitemOpen
  \bibfield  {author} {\bibinfo {author} {\bibfnamefont {J.~J.}\ \bibnamefont
  {Niemela}}, \bibinfo {author} {\bibfnamefont {L.}~\bibnamefont {Skrbek}},
  \bibinfo {author} {\bibfnamefont {K.~R.}\ \bibnamefont {Sreenivasan}}, \ and\
  \bibinfo {author} {\bibfnamefont {R.~J.}\ \bibnamefont {Donnelly}},\
  }\href@noop {} {\bibfield  {journal} {\bibinfo  {journal} {Nature}\ }\textbf
  {\bibinfo {volume} {404}},\ \bibinfo {pages} {837} (\bibinfo {year}
  {2000})}\BibitemShut {NoStop}%
\bibitem [{\citenamefont {Boffetta}\ \emph {et~al.}(2012)\citenamefont
  {Boffetta}, \citenamefont {De~Lillo}, \citenamefont {Mazzino},\ and\
  \citenamefont {Musacchio}}]{boffetta2012bolgiano}%
  \BibitemOpen
  \bibfield  {author} {\bibinfo {author} {\bibfnamefont {G.}~\bibnamefont
  {Boffetta}}, \bibinfo {author} {\bibfnamefont {F.}~\bibnamefont {De~Lillo}},
  \bibinfo {author} {\bibfnamefont {A.}~\bibnamefont {Mazzino}}, \ and\
  \bibinfo {author} {\bibfnamefont {S.}~\bibnamefont {Musacchio}},\ }\href@noop
  {} {\bibfield  {journal} {\bibinfo  {journal} {J. Fluid Mech.}\ }\textbf
  {\bibinfo {volume} {690}},\ \bibinfo {pages} {426} (\bibinfo {year}
  {2012})}\BibitemShut {NoStop}%
\bibitem [{\citenamefont {Aivalis}\ \emph {et~al.}(2002)\citenamefont
  {Aivalis}, \citenamefont {Sreenivasan}, \citenamefont {Tsuji}, \citenamefont
  {Klewicki},\ and\ \citenamefont {Biltoft}}]{aivalis2002temperature}%
  \BibitemOpen
  \bibfield  {author} {\bibinfo {author} {\bibfnamefont {K.~G.}\ \bibnamefont
  {Aivalis}}, \bibinfo {author} {\bibfnamefont {K.~R.}\ \bibnamefont
  {Sreenivasan}}, \bibinfo {author} {\bibfnamefont {Y.}~\bibnamefont {Tsuji}},
  \bibinfo {author} {\bibfnamefont {J.~C.}\ \bibnamefont {Klewicki}}, \ and\
  \bibinfo {author} {\bibfnamefont {C.~A.}\ \bibnamefont {Biltoft}},\
  }\href@noop {} {\bibfield  {journal} {\bibinfo  {journal} {Phys.\ Fluids}\
  }\textbf {\bibinfo {volume} {14}},\ \bibinfo {pages} {2439} (\bibinfo {year}
  {2002})}\BibitemShut {NoStop}%
\bibitem [{\citenamefont {Lovejoy}\ \emph {et~al.}(2007)\citenamefont
  {Lovejoy}, \citenamefont {Tuck}, \citenamefont {Hovde},\ and\ \citenamefont
  {Schertzer}}]{lovejoy2007isotropic}%
  \BibitemOpen
  \bibfield  {author} {\bibinfo {author} {\bibfnamefont {S.}~\bibnamefont
  {Lovejoy}}, \bibinfo {author} {\bibfnamefont {A.}~\bibnamefont {Tuck}},
  \bibinfo {author} {\bibfnamefont {S.}~\bibnamefont {Hovde}}, \ and\ \bibinfo
  {author} {\bibfnamefont {D.}~\bibnamefont {Schertzer}},\ }\href@noop {}
  {\bibfield  {journal} {\bibinfo  {journal} {Geophys. Res. Lett.}\ }\textbf
  {\bibinfo {volume} {34}},\ \bibinfo {pages} {L15802} (\bibinfo {year}
  {2007})}\BibitemShut {NoStop}%
\bibitem [{\citenamefont {Lovejoy}\ and\ \citenamefont
  {Schertzer}(2013)}]{lovejoy2013weather}%
  \BibitemOpen
  \bibfield  {author} {\bibinfo {author} {\bibfnamefont {S.}~\bibnamefont
  {Lovejoy}}\ and\ \bibinfo {author} {\bibfnamefont {D.}~\bibnamefont
  {Schertzer}},\ }\href@noop {} {\emph {\bibinfo {title} {The Weather and
  Climate: Emergent Laws and Multifractal Cascades}}}\ (\bibinfo  {publisher}
  {Cambridge University Press},\ \bibinfo {year} {2013})\ p.\ \bibinfo {pages}
  {505 pp.}\BibitemShut {Stop}%
\bibitem [{Note3()}]{Note3}%
  \BibitemOpen
  \bibinfo {note} {The text inside the parentheses are made by the authors of
  the present paper and not by \cite {monin1975}}\BibitemShut {NoStop}%
\bibitem [{Note4()}]{Note4}%
  \BibitemOpen
  \bibinfo {note} {A second-order central difference scheme (with non-uniform
  spacing) is used for the slope calculations.}\BibitemShut {Stop}%
\bibitem [{\citenamefont {Blackadar}(1957)}]{Blackadar1957}%
  \BibitemOpen
  \bibfield  {author} {\bibinfo {author} {\bibfnamefont {A.~K.}\ \bibnamefont
  {Blackadar}},\ }\href@noop {} {\bibfield  {journal} {\bibinfo  {journal}
  {Bull. Amer. Meteorol. Soc.}\ }\textbf {\bibinfo {volume} {38}},\ \bibinfo
  {pages} {283} (\bibinfo {year} {1957})}\BibitemShut {NoStop}%
\bibitem [{\citenamefont {Bonner}(1968)}]{bonn68}%
  \BibitemOpen
  \bibfield  {author} {\bibinfo {author} {\bibfnamefont {W.~D.}\ \bibnamefont
  {Bonner}},\ }\href@noop {} {\bibfield  {journal} {\bibinfo  {journal} {Mon.
  Wea. Rev.}\ }\textbf {\bibinfo {volume} {96}},\ \bibinfo {pages} {833}
  (\bibinfo {year} {1968})}\BibitemShut {NoStop}%
\bibitem [{\citenamefont {Sisterson}\ and\ \citenamefont
  {Frenzen}(1978)}]{sist78}%
  \BibitemOpen
  \bibfield  {author} {\bibinfo {author} {\bibfnamefont {D.~L.}\ \bibnamefont
  {Sisterson}}\ and\ \bibinfo {author} {\bibfnamefont {P.}~\bibnamefont
  {Frenzen}},\ }\href@noop {} {\bibfield  {journal} {\bibinfo  {journal}
  {Environ. Sci. Tech.}\ }\textbf {\bibinfo {volume} {12}},\ \bibinfo {pages}
  {218} (\bibinfo {year} {1978})}\BibitemShut {NoStop}%
\bibitem [{\citenamefont {Song}\ \emph {et~al.}(2005)\citenamefont {Song},
  \citenamefont {Liao}, \citenamefont {Coulter},\ and\ \citenamefont
  {Lesht}}]{Songetal2005}%
  \BibitemOpen
  \bibfield  {author} {\bibinfo {author} {\bibfnamefont {J.}~\bibnamefont
  {Song}}, \bibinfo {author} {\bibfnamefont {K.}~\bibnamefont {Liao}}, \bibinfo
  {author} {\bibfnamefont {R.}~\bibnamefont {Coulter}}, \ and\ \bibinfo
  {author} {\bibfnamefont {B.}~\bibnamefont {Lesht}},\ }\href@noop {}
  {\bibfield  {journal} {\bibinfo  {journal} {J. Appl. Meteorol.}\ }\textbf
  {\bibinfo {volume} {44}},\ \bibinfo {pages} {1593} (\bibinfo {year}
  {2005})}\BibitemShut {NoStop}%
\bibitem [{\citenamefont {Rife}\ \emph {et~al.}(2010)\citenamefont {Rife},
  \citenamefont {Pinto}, \citenamefont {Monaghan}, \citenamefont {Davis},\ and\
  \citenamefont {Hannan}}]{rife10}%
  \BibitemOpen
  \bibfield  {author} {\bibinfo {author} {\bibfnamefont {D.~L.}\ \bibnamefont
  {Rife}}, \bibinfo {author} {\bibfnamefont {J.~O.}\ \bibnamefont {Pinto}},
  \bibinfo {author} {\bibfnamefont {A.~J.}\ \bibnamefont {Monaghan}}, \bibinfo
  {author} {\bibfnamefont {C.~A.}\ \bibnamefont {Davis}}, \ and\ \bibinfo
  {author} {\bibfnamefont {J.~R.}\ \bibnamefont {Hannan}},\ }\href@noop {}
  {\bibfield  {journal} {\bibinfo  {journal} {J. Clim.}\ }\textbf {\bibinfo
  {volume} {23}},\ \bibinfo {pages} {5041} (\bibinfo {year}
  {2010})}\BibitemShut {NoStop}%
\bibitem [{\citenamefont {Storm}\ \emph {et~al.}(2009)\citenamefont {Storm},
  \citenamefont {Dudhia}, \citenamefont {Basu}, \citenamefont {Swift},\ and\
  \citenamefont {Giammanco}}]{storm2009evaluation}%
  \BibitemOpen
  \bibfield  {author} {\bibinfo {author} {\bibfnamefont {B.}~\bibnamefont
  {Storm}}, \bibinfo {author} {\bibfnamefont {J.}~\bibnamefont {Dudhia}},
  \bibinfo {author} {\bibfnamefont {S.}~\bibnamefont {Basu}}, \bibinfo {author}
  {\bibfnamefont {A.}~\bibnamefont {Swift}}, \ and\ \bibinfo {author}
  {\bibfnamefont {I.}~\bibnamefont {Giammanco}},\ }\href@noop {} {\bibfield
  {journal} {\bibinfo  {journal} {Wind Energy}\ }\textbf {\bibinfo {volume}
  {12}},\ \bibinfo {pages} {81} (\bibinfo {year} {2009})}\BibitemShut {NoStop}%
\bibitem [{\citenamefont {Storm}\ and\ \citenamefont {Basu}(2010)}]{stor10}%
  \BibitemOpen
  \bibfield  {author} {\bibinfo {author} {\bibfnamefont {B.}~\bibnamefont
  {Storm}}\ and\ \bibinfo {author} {\bibfnamefont {S.}~\bibnamefont {Basu}},\
  }\href@noop {} {\bibfield  {journal} {\bibinfo  {journal} {Energies}\
  }\textbf {\bibinfo {volume} {3}},\ \bibinfo {pages} {258} (\bibinfo {year}
  {2010})}\BibitemShut {NoStop}%
\bibitem [{\citenamefont {Wilczak}\ \emph {et~al.}(2014)\citenamefont
  {Wilczak}, \citenamefont {Finley}, \citenamefont {Freedman}, \citenamefont
  {Cline}, \citenamefont {Bianco}, \citenamefont {Olson}, \citenamefont
  {Djalalova}, \citenamefont {Sheridan}, \citenamefont {Ahlstrom},
  \citenamefont {Manobianco} \emph {et~al.}}]{wilczak2014wind}%
  \BibitemOpen
  \bibfield  {author} {\bibinfo {author} {\bibfnamefont {J.}~\bibnamefont
  {Wilczak}}, \bibinfo {author} {\bibfnamefont {C.}~\bibnamefont {Finley}},
  \bibinfo {author} {\bibfnamefont {J.}~\bibnamefont {Freedman}}, \bibinfo
  {author} {\bibfnamefont {J.}~\bibnamefont {Cline}}, \bibinfo {author}
  {\bibfnamefont {L.}~\bibnamefont {Bianco}}, \bibinfo {author} {\bibfnamefont
  {J.}~\bibnamefont {Olson}}, \bibinfo {author} {\bibfnamefont
  {I.}~\bibnamefont {Djalalova}}, \bibinfo {author} {\bibfnamefont
  {L.}~\bibnamefont {Sheridan}}, \bibinfo {author} {\bibfnamefont
  {M.}~\bibnamefont {Ahlstrom}}, \bibinfo {author} {\bibfnamefont
  {J.}~\bibnamefont {Manobianco}},  \emph {et~al.},\ }\href@noop {} {\bibfield
  {journal} {\bibinfo  {journal} {Bull. Amer. Meteorol. Soc.}\ } (\bibinfo
  {year} {2014})},\ \bibinfo {note} {doi:
  http://dx.doi.org/10.1175/BAMS-D-14-00107.1}\BibitemShut {NoStop}%
\bibitem [{\citenamefont {Van~de Wiel}\ \emph {et~al.}(2010)\citenamefont
  {Van~de Wiel}, \citenamefont {Moene}, \citenamefont {Steeneveld},
  \citenamefont {Baas}, \citenamefont {Bosveld},\ and\ \citenamefont
  {Holtslag}}]{van2010conceptual}%
  \BibitemOpen
  \bibfield  {author} {\bibinfo {author} {\bibfnamefont {B.~J.~H.}\
  \bibnamefont {Van~de Wiel}}, \bibinfo {author} {\bibfnamefont {A.~F.}\
  \bibnamefont {Moene}}, \bibinfo {author} {\bibfnamefont {G.-J.}\ \bibnamefont
  {Steeneveld}}, \bibinfo {author} {\bibfnamefont {P.}~\bibnamefont {Baas}},
  \bibinfo {author} {\bibfnamefont {F.~C.}\ \bibnamefont {Bosveld}}, \ and\
  \bibinfo {author} {\bibfnamefont {A.~A.~M.}\ \bibnamefont {Holtslag}},\
  }\href@noop {} {\bibfield  {journal} {\bibinfo  {journal} {J. Atmos. Sci.}\
  }\textbf {\bibinfo {volume} {67}},\ \bibinfo {pages} {2679} (\bibinfo {year}
  {2010})}\BibitemShut {NoStop}%
\bibitem [{\citenamefont {Parish}\ and\ \citenamefont
  {Oolman}(2010)}]{parish2010role}%
  \BibitemOpen
  \bibfield  {author} {\bibinfo {author} {\bibfnamefont {T.~R.}\ \bibnamefont
  {Parish}}\ and\ \bibinfo {author} {\bibfnamefont {L.~D.}\ \bibnamefont
  {Oolman}},\ }\href@noop {} {\bibfield  {journal} {\bibinfo  {journal} {J.
  Atmos. Sci.}\ }\textbf {\bibinfo {volume} {67}},\ \bibinfo {pages} {2690}
  (\bibinfo {year} {2010})}\BibitemShut {NoStop}%
\bibitem [{\citenamefont {Ruiz-Chavarria}\ \emph {et~al.}(2000)\citenamefont
  {Ruiz-Chavarria}, \citenamefont {Ciliberto}, \citenamefont {Baudet},\ and\
  \citenamefont {L{\'e}v{\^e}que}}]{ruiz2000scaling}%
  \BibitemOpen
  \bibfield  {author} {\bibinfo {author} {\bibfnamefont {G.}~\bibnamefont
  {Ruiz-Chavarria}}, \bibinfo {author} {\bibfnamefont {S.}~\bibnamefont
  {Ciliberto}}, \bibinfo {author} {\bibfnamefont {C.}~\bibnamefont {Baudet}}, \
  and\ \bibinfo {author} {\bibfnamefont {E.}~\bibnamefont {L{\'e}v{\^e}que}},\
  }\href@noop {} {\bibfield  {journal} {\bibinfo  {journal} {Physica D}\
  }\textbf {\bibinfo {volume} {141}},\ \bibinfo {pages} {183} (\bibinfo {year}
  {2000})}\BibitemShut {NoStop}%
\bibitem [{\citenamefont {Benzi}\ \emph {et~al.}(1993)\citenamefont {Benzi},
  \citenamefont {Ciliberto}, \citenamefont {Tripiccione}, \citenamefont
  {Baudet}, \citenamefont {Massaioli},\ and\ \citenamefont
  {Succi}}]{benzi1993extended}%
  \BibitemOpen
  \bibfield  {author} {\bibinfo {author} {\bibfnamefont {R.}~\bibnamefont
  {Benzi}}, \bibinfo {author} {\bibfnamefont {S.}~\bibnamefont {Ciliberto}},
  \bibinfo {author} {\bibfnamefont {R.}~\bibnamefont {Tripiccione}}, \bibinfo
  {author} {\bibfnamefont {C.}~\bibnamefont {Baudet}}, \bibinfo {author}
  {\bibfnamefont {F.}~\bibnamefont {Massaioli}}, \ and\ \bibinfo {author}
  {\bibfnamefont {S.}~\bibnamefont {Succi}},\ }\href@noop {} {\bibfield
  {journal} {\bibinfo  {journal} {Phys.\ Rev.\ E}\ }\textbf {\bibinfo {volume}
  {48}},\ \bibinfo {pages} {R29} (\bibinfo {year} {1993})}\BibitemShut
  {NoStop}%
\bibitem [{\citenamefont {Arneodo}\ \emph {et~al.}(1996)\citenamefont
  {Arneodo}, \citenamefont {Baudet}, \citenamefont {Belin}, \citenamefont
  {Benzi}, \citenamefont {Castaing}, \citenamefont {Chabaud}, \citenamefont
  {Chavarria}, \citenamefont {Ciliberto}, \citenamefont {Camussi},
  \citenamefont {Chilla} \emph {et~al.}}]{arneodo1996structure}%
  \BibitemOpen
  \bibfield  {author} {\bibinfo {author} {\bibfnamefont {A.}~\bibnamefont
  {Arneodo}}, \bibinfo {author} {\bibfnamefont {C.}~\bibnamefont {Baudet}},
  \bibinfo {author} {\bibfnamefont {F.}~\bibnamefont {Belin}}, \bibinfo
  {author} {\bibfnamefont {R.}~\bibnamefont {Benzi}}, \bibinfo {author}
  {\bibfnamefont {B.}~\bibnamefont {Castaing}}, \bibinfo {author}
  {\bibfnamefont {B.}~\bibnamefont {Chabaud}}, \bibinfo {author} {\bibfnamefont
  {R.}~\bibnamefont {Chavarria}}, \bibinfo {author} {\bibfnamefont
  {S.}~\bibnamefont {Ciliberto}}, \bibinfo {author} {\bibfnamefont
  {R.}~\bibnamefont {Camussi}}, \bibinfo {author} {\bibfnamefont
  {F.}~\bibnamefont {Chilla}},  \emph {et~al.},\ }\href@noop {} {\bibfield
  {journal} {\bibinfo  {journal} {Europhys. Lett.}\ }\textbf {\bibinfo {volume}
  {34}},\ \bibinfo {pages} {411} (\bibinfo {year} {1996})}\BibitemShut
  {NoStop}%
\bibitem [{\citenamefont {Basu}\ \emph {et~al.}(2006)\citenamefont {Basu},
  \citenamefont {Port{\'e}-Agel}, \citenamefont {Foufoula-Georgiou},
  \citenamefont {Vinuesa},\ and\ \citenamefont {Pahlow}}]{basu2006revisiting}%
  \BibitemOpen
  \bibfield  {author} {\bibinfo {author} {\bibfnamefont {S.}~\bibnamefont
  {Basu}}, \bibinfo {author} {\bibfnamefont {F.}~\bibnamefont
  {Port{\'e}-Agel}}, \bibinfo {author} {\bibfnamefont {E.}~\bibnamefont
  {Foufoula-Georgiou}}, \bibinfo {author} {\bibfnamefont {J.-F.}\ \bibnamefont
  {Vinuesa}}, \ and\ \bibinfo {author} {\bibfnamefont {M.}~\bibnamefont
  {Pahlow}},\ }\href@noop {} {\bibfield  {journal} {\bibinfo  {journal}
  {Boundary-Layer Meteorology}\ }\textbf {\bibinfo {volume} {119}},\ \bibinfo
  {pages} {473} (\bibinfo {year} {2006})}\BibitemShut {NoStop}%
\end{thebibliography}
\providecommand{\noopsort}[1]{}\providecommand{\singleletter}[1]{#1}%

\end{document}